\documentclass[12pt,psfig,times,mathptm]{article}

\usepackage{amsmath}
\usepackage{array}
\usepackage{hyperref}
\usepackage{caption}
\usepackage{subcaption}

\usepackage{graphicx}
\usepackage{makecell}
\usepackage{float} 
\usepackage{cancel}

\usepackage{multirow}

\begin{document}
\baselineskip 0.6cm

\newcommand{\beq}{\begin{equation}}
\newcommand{\eeq}{\end{equation}}
\newcommand{\bea}{\begin{eqnarray}}
\newcommand{\eea}{\end{eqnarray}}
\newcommand{\beas}{\begin{eqnarray*}}
\newcommand{\eeas}{\end{eqnarray*}}
\newcommand{\tri}{\triangleright}
\newcommand{\range}{{\rm range}}
\newcommand{\Ree}{{\rm Re }}
\newcommand{\Imm}{{\rm Im }}
\newcommand{\diag}{{\rm diag}}
\newcommand{\sign}{{\rm sign}}
\newcommand{\tr}{{\rm tr}}
\newcommand{\rank}{{\rm rank}}
\newcommand{\bp}{\bigskip}
\newcommand{\mdp}{\medskip}
\newcommand{\slp}{\smallskip}
\newcommand{\Rw}{\Rightarrow}
\newcommand{\ts}{& \hspace{-0.1in}}
\newcommand{\nn}{\nonumber}
\newtheorem{exa}{Example}[section]
\newtheorem{thm}{Theorem}[section]
\newtheorem{lem}{Lemma}[section]
\newtheorem{prop}{Proposition}[section]
\newtheorem{fact}{Fact}[section]
\newtheorem{cor}{Corollary}[section]
\newtheorem{defn}{Definition}[section]
\newtheorem{rem}{Remark}[section]
\renewcommand{\theequation}{\thesection.\arabic{equation}}
\title{Koopman System Approximation Based Optimal Control of Multiple
  Robots---Part II: Simulations and Evaluations}
\author{{\it Qianhong Zhao} and {\it Gang Tao} \\
Department of Electrical and Computer Engineering\\
University of Virginia\\
Charlottesville, VA 22904\\ \\
}
\date{}
\maketitle
\begin{abstract}

This report presents the results of a simulation study of the linear
model and bilinear model approximations of the Koopman system model
 of the nonlinear utility functions in optimal control of
a 3-robot system. In such a control problem, the nonlinear utility
functions are maximized to achieve the control objective of moving
the robots to their target positions and avoiding collisions.
With the linear and bilinear model approximations of the utility
functions, the optimal control problem is solved, based on the
approximate model state variables rather than the original nonlinear
utility functions. This transforms the original nonlinear game theory problem to a linear optimization problem. This report studies both the centralized and decentralized implementations of the approximation model based control signals for the 3-robot system control problem.  

\medskip
The simulation results show that the \textit{posteriori} estimation error of the linear
approximation model has a dramatic tendency of periodic increase in values, which indicates that a linear
model cannot approximate the nonlinear utility function well, and
is not suitable for solving the optimal control problem (as verified
by simulation results: the control signals from the linear model based
optimization cannot make the robots move to their targets).

\medskip
The simulation results also show that the \textit{posteriori} estimation error of the
bilinear approximation model shows the error of the bilinear approximation model is much smaller than the error of the linear approximation model. Although the tendency of periodic increase appears again, the maximum value of the \textit{posteriori} estimation error of the bilinear approximation model is several thousand times less than the linear approxiamtion model. This indicates
that the bilinear model has more capacity to approximate the nonlinear
utility functions. Both the centralized and decentralized bilinear approximation model based control
signals can achieve the control objective of moving the robots to
their target positions within a short time. Based on the analysis
 of the simulation time, the bilinear model based optimal control
 solution is fast enough for real-time control implementation.
   
\end{abstract}

\begin{quote}
\end{quote}

\setcounter{equation}{0}
\section{Introduction}
\label{Sec_Intro}
Approximation of a nonlinear dynamic system by a linear dynamic system
is a useful approach for the control of nonlinear systems. Recently,
Koopman operator theory and Koopman transformation techniques have
been employed to further develop such an approach. The basic idea of
the Koopman system approach is, for a given dynamic system, to first
introduce a set of measured signals of interest (called the
observables or Koopman variables), then generate a set of dynamic
equations using Koopman operator theory and Koopman transformation
techniques, and finally form a linear system structure and calculate
its parameters using some minimization techniques. 
More details about this approach can be found in \cite{Williams2015ADA},\cite{km18},\cite{Yeung2019LearningDN},\cite{Bevanda2021KoopmanOD},\cite{mazouchi2021finitetime},\cite{t21}.

In our study, we consider a 3-robot system with some nonlinear utility
functions defined from an optimal control problem, select the utility
functions as the Koopman variables to define a Koopman dynamic system,
form a linear or a bilinear system structure to approximate the nonlinear system
(the augmented robot system), propose an adaptive algorithm to
estimate (identify) the unknown parameters of the linear model and bilinear model online and the optimal control designs based on the estimated linear and bilinear models.

In this report, we summarize the results of our study: not only the linear model identification algorithm and the linear model based optimal control design, but also the bilinear model identification and the bilinear model based optimal control design are simulated on the 3-robot system. However, the simulation results show that only the bilinear approximation model is capable to estimate the nonlinear utility function precisely for both centralized and decentralized control designs to the control problem of 3-robot system. The robot trajectories and program running time of simulation results prove the ability of the control design based on the bilinear model is fast enough to solve the real problem.

This report has the following contents. In Section \ref{Sec_MultiSys}, the multi-robot simulation system is introduced, including the linear motion equations of the
robots with some nonlinear utility functions from an optimal control
problem, the defined Koopman system state variables, and the optimal control design idea based on the linear model or the bilinear model. In Section \ref{Sec_ParaEst}, an adaptive
(iterative) parameter estimation algorithm (for linear and bilinear model identification). Section \ref{Sec_DecThe} demonstrates the decentralized implementation of the bilinear approximation model based control design for the 3-robot system. 
In Section \ref{Sec_Simu}, the simulation results are presented,
including the parameter estimation (system identification) results for the 3-robot system, the linear model based on control design for the 3-robot system and the bilinear model based on control design for the 3-robot system. In Section 5, the concluding remarks of the simulation results are summarized.

\setcounter{equation}{0}

\section{Multi-Robot System with Nonlinear Utility Functions}
\label{Sec_MultiSys}
In this section, the multi-robot system models with nonlinear utility functions are detailed. Multi-robot dynamic system models, nonlinear utility functions, linear Koopman system models, and the optimal control problem are introduced in subsections 2.1-2.4 correspondingly. 

\subsection{Multi-Robot System Models}
In this subsection, the multi-robot dynamic system model is introduced. 

\subsubsection{System with 3 Robots}
 The 3-robot system is applied to examine the performance of the estimated model based control designs. The 3-robot system models 
are described as
\bea
v_{i}(k+1) \ts = \ts v_{i}(k) + a_{i}(k)\, \Delta t \nn\\
r_{i}(k+1) \ts = \ts r_{i}(k) + v_{i}(k)\, \Delta t + 0.5\,
a_{i}(k)\,(\Delta t)^2,
\label{3r_robots_indiv}
\eea
where $r_i(k)=\left ( 
\begin{matrix}
x_i(k)\\
y_i(k)
\end{matrix}
\right)$, 
$ v_i(k)=\left(
\begin{matrix}
v_i^x(k)\\
v_i^y(k)
\end{matrix}\right)$, and $a_i(k)=\left(
\begin{matrix}
a_i^x(k)\\
a_i^y(k)
\end{matrix}\right),i = 1,2,3$.

With the total state vector
\beq
X(k) = [x_{1}(k), y_{1}(k), \ldots, x_{3}(k), y_{3}(k), v_{1}^x(k),
  v_{1}^y(k), \ldots, v_{3}^x(k), v_{3}^y(k)]^T \in R^{12},
\label{3r_z1}
\eeq
we can express (\ref{3r_robots_indiv}) as
\beq
X(k+1) = A X(k) + B U(k),
\label{3r_z1_update}
\eeq
where the control vector $U(k)$ is
\beq
U(k) = [a_{1}^x(k), a_{1}^y(k), a_{2}^x(k), a_{2}^y(k), 
  a_{3}^x(k), a_{3}^y(k)]^T \in R^{6},
\eeq
and $A \in R^{12 \times 12}$ and $B \in R^{12 \times 6}$ are some
matrices as the controlled system matrices which can be directly
obtained from the system models (\ref{3r_robots_indiv}):
\beq
A = \left [
\begin{matrix}
I_{6\times6}& \Delta t\, I_{6\times6}\\
0_{6\times6}& I_{6\times6}\\
\end{matrix}
\right],B = \left [
\begin{matrix}
0.5(\Delta t)^2\,I_{6\times 6}\\
\Delta t\,I_{6\times 6}
\end{matrix}
\right] ,
\eeq
where $\Delta t$ is the sampling period. 
\subsection{Nonlinear Utility Functions}
In \cite{wetal20}, some nonlinear utility functions based on game theory are proposed to make the robots arrive at their targets without collisions. The utility functions of the $3$-robot system have the form
\beq
u_{i}(k) = \sum_{j=1}^6 \omega_i^{(j)} \phi_{i}^{(j)}(X(k)), i = 1,2,3,
\label{utility2}
\eeq
where $X(k)$ in (\ref{3r_z1}) depends on $s_{l}(k) = \{r_{l}(k),
v_{l}(k)\}$, $(l = 1,2,3)$, $\omega_i^{(j)}$ are 
constants, and $\phi_{i}^{(j)}(X(k))$ are given in \cite{wetal20}, in
terms of $s_{l}(k) = \{r_{l}(k), v_{l}(k)\}$:
\beq
\phi_{i}^{(1)}(X(k)) = \frac{r^T_{i*}(k)v_{i}(k)}{||r_{i*}(k)||_2||v_{i}(k)||_2}, 
\label{phi_c1}
\eeq
where $r_{i*}(k) = (x_{i}(k)-x_{i*}, \,y_{i}(k)-y_{i*}) $ expresses the difference between current position $(x_{i}(k), \,y_{i}(k))$ and the desired position $(x_{i*}, \,y_{i*})$,
\beq
\phi_{i}^{(2)}(X(k)) = 1-\exp(-(\frac{|v_{i}(k)|-v^*(r_{i*}(k))}{0.8v^*(r_{i*}(k))})^2), 
\eeq
where $v^*(r_{i*}(k)) = \frac{4}{1+\exp(-10(|r_{i*}(k)|-0.2))}$,
\beq
\phi_{i}^{(3)}(X(k)) = [\exp(-(\frac{|r_{i*}(k)|}{4.0})^2)] \times [\exp(-(\frac{|r_{i*}(k)|}{6.0})^2)],
\eeq
\beq
\phi_{i}^{(4)}(X(k)) = [\exp(-(\frac{|r_{i*}(k)|}{0.05})^2)]\times [\exp(-(\frac{|v_{i}(k)|}{0.05})^2)],
\eeq
\beq
\phi_{i}^{(5)}(X(k)) = \ln[1+6\times\exp(-40\times d_{iw}(k))],
\eeq
where $d_{iw}(k)=|r_{iw}(k)|-R_r=R_w-||(x_{i}(k)-x_c,y_{i}(k)-y_c)||_2-R_r$ is the shortest distance to the wall which can be computed by the circular wall radius $R_w$, the Robot $i$ current position $(x_i,y_i)$, the center of the circular wall $(x_c,y_c)$ and the robot radius $R_r$.
\beq
\phi_{i}^{(6)}(X(k)) = \ln[1+10\times\exp(-20\times d_{ij}(k))],
\label{phi_c6}
\eeq
where $d_{ij}(k)=|r_{ij}(k)|-2R_r= \min ||(x_{i}(k)-x_{j}(k),y_{i}(k)-y_{j}(k))||_2-R_r , \; i = 1,2,3,\;i\neq j,$ is the shortest distance between Robot$_i$ and other robots.

The Koopman variables are defined as
\beq
z_{i,j}(k) = \psi_{i,j}(X(k)) =  \phi_{i}^{(j)}(X(k)),
\eeq
for $i = 1,2,3,\;j=1,2,\ldots, 6$, and the corresponding
Koopman equations are
\beq
\psi_{i,j}(X(k+1)) =  \psi_{i,j}(A X(k) + B U(k)) = 
\phi_{i}^{(j)}(A X(k) + B U(k)).
\label{psiij}
\eeq

Our goal of study is to express $\phi_{i}^{(j)}(A X(k) + B U(k))$ in terms of 
$\psi_{i,j}(X(k))$, $X(k)$ and $U(k)$, i.e., $z_{i,j}(k+1) =
\psi_{i,j}(X(k+1))$ in terms of $z_{i,j}(k) =
\psi_{i,j}(X(k))$ and $X(k)$ and $U(k)$. 
\subsection{Koopman System Models}

This subsection presents the detailed Koopman system models to acquire a linear representation for the multi-robot system with nonlinear utility functions. 
\subsubsection{Koopman System Variables}

It is necessary to select some Koopman variables to apply the Koopman transformation techniques. 

\bigskip
\textbf{Natural state variables}. We first set the original system state
variables of the $3$-robot system in (\ref{3r_z1}) as
\beq
z_i(k) = \psi_i(X(k)) = X_i(k),\;i=1,2,\ldots, 12,
\label{zig1}
\eeq
where $X_i(k)$ are the components of $X(k)$:
\bea
X(k) \ts = \ts [X_1(k), X_2(k), \ldots, X_{12}(k)]^T\;  \nn \\
\ts \stackrel{\triangle} = \ts 
[x_{1}(k), y_{1}(k), \ldots, x_{3}(k), y_{3}(k), v_{1}^x(k),
  v_{1}^y(k), \ldots, v_{3}^x(k), v_{3}^y(k)]^T \in R^{12},
\eea

For this choice of $z_i(k)=\psi_{i}(X(k))=X_i(k)$, the Koopman equations: $\psi_{i}(f(x(k),u(k)))=\psi_{i}(x(k+1))=z_i(k+1)$ for $x(k+1)=f(x(k),u(k))$, applied to $X(k+1)=AX(k)+BU(k)$, become 
\begin{equation}
    \psi_{i}(AX(k)+BU(k))=\psi_{i}(X(k+1))\stackrel{\psi_{i}(X(k))=X_i(k)}{\Rightarrow}(AX(k)+BU(k))_i=X_i(k+1)=z_i(k+1),
\end{equation}
where $(AX(k)+BU(k))_i$ denotes the $i$th component of $AX(k)+BU(k)$, or in the vector form,
\begin{equation}
    AX(k)+BU(k)=X(k+1)\Leftrightarrow Az_{(1)}(k)+BU(k)=z_{(1)}(k+1),
    \label{Koopmansubsys1}
\end{equation}
where 
\beq
z_{(1)}(k) = [z_1(k),\cdots,z_{12}(k)]^T=X(k)\in R^{12},
\label{z1_def}
\eeq
that is, the Koopman subsystem (\ref{Koopmansubsys1}) for $z_{(1)}=[z_1(k),\cdots,z_{12}(k)]$ is the original system (\ref{3r_z1_update}) itself which is already linear.

\bigskip
\textbf{Utility function components}. We then introduce the nonlinear utility function components based on the introduced Koopman variables in (\ref{zig1}):
\bea
\ts \ts z_{12+6(i-1)+j}(k) = \psi_{12+6(i-1)+j}(X(k)) \nn\\
\ts = \ts z_{i,j}(k) =
\psi_{i,j}(X(k)) = \phi_{i}^{(j)}(X(k)),\;i=1,2,3,\;j=1,2,\ldots, 6, 
\label{zig2}
\eea
(from $z_{13}(k)$ to $z_{30}(k)$),
 for $\phi_{i}^{(j)}(X(k))$ in the utility function expression
(\ref{utility2}): 
\beq
u_{i}(k) = \sum_{j=1}^6 \omega_i^{(j)}
\phi_{i}^{(j)}(X(k)),
\label{uik}
\eeq
to transform the nonlinear utility functions to a linear form:
\beq
u_{i}(k) = \sum_{j=1}^6 \omega_i^{(j)} z_{12+6(i-1)+j}(k),
\eeq
in terms of the Koopman system variables $z_{12+6(i-1)+j}(k)$,
$i=1,2,3,\;j=1,2,\ldots, 6$.

For the utility functions based definitions of Koopman variables $z_{12+6(i-1)+j}(k)$ calculated from (\ref{zig2}), the Koopman equations:
\beq
\psi_{i}(f(x(k),u(k)))=\psi_{i}(x(k+1))=z_i(k+1) ,
\eeq
for $x(k+1) = f(x(k),u(k))$, applied to $X(k+1)=AX(k)+BU(k)$, become 
\beq
\begin{split}
    &\psi_{m}(AX(k)+BU(k))=\psi_{m}(X(k+1))\\
    \stackrel{z_i(k)=\psi_{m}(X(k))=\phi_i^{(j)}(X(k))}{\Rightarrow} \;&\phi_i^{(j)}(AX(k)+BU(k))= \phi_i^{(j)}(X(k+1))=z_m(k+1),
\end{split}
\eeq
for $m=12+6(i-1)+j$ (with $i=1,2,3$, $j=1,2,\cdots,6$).

Similar to $z_{(1)}(k)$ in (\ref{z1_def}), we can define 
\beq
z_{(2)}(k)= [z_{13}(k),\cdots,z_{30}(k)]^T\in R^{18},
\label{z2_def}
\eeq
as the second group of Koopman variables. The total system state vector is
\beq
z(k) = [(z_{(1)}(k))^T, (z_{(2)}(k))^T]^T \in R^{30}.
\eeq
In a compact form, we have 
\begin{equation}
    z_{(2)}(k) = [( z_{(2)1}(k))^T,( z_{(2)2}(k))^T,( z_{(2)3}(k))^T]^T\in R^{18},
\end{equation}
where \begin{equation}
     z_{(2)1}(k) = [z_{13}(k),\cdots, z_{18}(k)]^T\in R^6,
\end{equation}
\begin{equation}
     z_{(2)2}(k) = [z_{19}(k),\cdots, z_{24}(k)]^T\in R^6,
\end{equation}
\begin{equation}
     z_{(2)3}(k) = [z_{25}(k),\cdots, z_{30}(k)]^T\in R^6.
\end{equation}

\subsubsection{Nonlinear Koopman System Model}
According to equations (\ref{phi_c1})-(\ref{phi_c6}) and (\ref{zig2}), we have the nonlinear Koopman system model $\psi_{i}(f(x(k),u(k)))=\psi_{i}(x(k+1))=z_i(k+1)$ for $x(k+1)=f(x(k),u(k))$ can be expressed as $z(k+1)=\xi(z(k),u(k))$, for some vector function $\xi(\cdot, \cdot)$. This implies, when applied to $\phi_{i}^{(j)}(AX(k)+BU(k))=\phi_{i}^{(j)}(X(k+1))$, 
\beq
z_{(2)}(k+1)=\xi_{(2)}(z_{(2)}(k),U(k))
\label{nonlinear_zk+1}
\eeq
for some vector function $\xi_{(2)}(\cdot,\cdot)\in R^{18}$, where $z_{(2)}(k)$ is defined in (\ref{z2_def}):
\begin{equation}
    z_{(2)}(k)=\psi_{(2)}(X(k))=[\psi_{13}(X(k)),\cdots\psi_{30}(X(k))]^T.
\end{equation}
For linear system identification, our goal is to express or approximate $\xi_{(2)}(z_{(2)}(k),U(k))$ by some linear functions of $z_{(2)}(k)$ (plus $z_{(1)}(k)$ and likely more Koopman variables to be introduced, to reduce the approximation error to make the linear model practically useful) and $U(k)$.

\subsubsection{Linear Koopman System Model and Parametrization}
This part introduces the linear approximation of the multi-robot system with nonlinear utility functions including the linear model and the corresponding parametrization. 

\bigskip
{\bf Linear system model for nonlinear utility functions.} With the Koopman state vector for $3$-robot system
\beq
z(k) = [(z_{(1)}(k))^T, (z_{(2)}(k))^T]^T \in R^{30},
\label{z(k)_l}
\eeq
where
\beq
z_{(1)}(k) = [z_1(k), \ldots, z_{12}(k)]^T = X(k) \in R^{12},
\label{z(1)_l}
\eeq
\beq
z_{(2)}(k) = [z_{13}(k), \ldots, z_{30}(k)]^T \in R^{18},
\label{z(2)_l}
\eeq
our interest is to find some matrices $A_z\in R^{30\times 30}$ and $B_z\in R^{30\times 6}$ such that
\begin{equation}
    z(k+1) = A_zz(k) + B_zU(k)+\eta_z(k),
\end{equation}
for some minimized error vector $\eta_{z}(k) \in R^{30}$. From (\ref{Koopmansubsys1}), we can see that $A_z$, $B_z$, and $\eta_z(k)$ have the following structures:

\begin{equation}
    \begin{split}
        A_z = \left[ \begin{array}{cc}
            A &0_{12\times 18}  \\
            A_{21} & A_{22} 
        \end{array}\right], B_z = \left[ \begin{array}{c}
            B \\
           B_2
        \end{array}\right], \eta_z(k) = \left[ \begin{array}{c}
            0_{12\times1} \\
           \eta_{(2)}(k)
        \end{array}\right],
    \end{split}
\end{equation}
where $A_{21} \in R^{18 \times 12}$,
$A_{22} \in R^{18 \times 18}$ and $B_2 \in R^{18 \times 6}$ are the parameter matrices to be calculated to minimize the error vector $\eta_{(2)}(k)\in R^{18}$. In other words, our task is to find the matrices
$A_{21} \in R^{18 \times 12}$,
$A_{22} \in R^{18 \times 18}$ and $B_2 \in R^{18 \times 6}$ to
express
\beq
z_{(2)}(k+1) = A_{21} z_{(1)}(k) + A_{22} z_{(2)}(k) + B_2  U(k) +
\eta_{(2)}(k),
\label{z(2)(k+1)}
\eeq
for some minimized error vector $\eta_{(2)}(k) \in R^{18}$.

\bigskip
{\bf Parametrization}. To develop an adaptive algorithm for
generating the  iterative (online) estimates of $A_{21} \in R^{18 \times 12}$, $A_{22} \in R^{18 \times 18}$ and $B_2 \in R^{18 \times
  6}$, we define 
\beq
y(k) = z_{(2)}(k+1) \in R^{18}
\label{y(k)(z_2)_l}
\eeq
\beq
\zeta(k) = [z_{(1)}^T(k), z_{(2)}^T(k), U^T(k)]^T \in R^{36}
\label{zeta}
\eeq
\beq
\Theta^T = [A_{21}, A_{22}, B_2] \in R^{18 \times 36}
\eeq
and rewrite (\ref{z(2)(k+1)}) as
\beq
y(k) = \Theta^T \zeta(k) + \eta_{(2)}(k),\;
\Theta^T = \left[\begin{array}{c}
\theta_1^T\\
\vdots\\
\theta_{18}^T
\end{array}
\right].
\label{y(k)_linear_paraed}
\eeq
With the estimates
\beq
\hat{\Theta}(k) = [\hat{A}_{21}(k), \hat{A}_{22}(k), \hat{B}_2(k)]^T
\label{theta_hat_linear}
\eeq of the above unknown parameter matrix $\Theta$, the approximated linear model for the nonlinear utility function (\ref{nonlinear_zk+1}) is 
\beq
\hat{z}_{(2)}(k+1) = \hat{A}_{21} {z}_{(1)}(k) + \hat{A}_{22} {z}_{(2)}(k) + \hat{B}_2  U(k),
\label{z(2)(k+1)appr_linear}
\eeq

\subsubsection{Bilinear Koopman System Model and Parametrization}

This subsection introduces the bilinear approximation of the $3$-robot system with nonlinear utility functions including the bilinear model and the corresponding parametrization. 

\bigskip
\par 
\textbf{Bilinear Model for Nonlinear Utility Functions.}
Since the simulation results demonstrate that the linear approximation Koopman model \cite{km18} (Korda and Mezic 2018) cannot solve nonlinear MPC problem, a bilinear model, an expansion of the linear Koopman model, is introduced in \cite{t22} and \cite{tzh22a}. The bilinear model of the form 
\begin{equation}
    z_{(2),i}(k+1) = f_i(x_0)+(\nabla f_i(x_0))^T(x-x_0)+\frac{1}{2}(x-x_0)^T H_{f_{i}}(x_0)(x-x_0),
\end{equation}
is proposed and estimated for its parameter matrices $f_i$, $\nabla f_i$ and $H_{f_i}$, where
\beq
x = [z^T_{(1)},z^T_{(2)},U^T]^T
\eeq
\beq
x_0 = [z^T_{(1)0},z^T_{(2)0},U_0^T]^T, U_0 = 0_{1\times6},
\label{x0}
\eeq
\beq
H_{f_{i}}(x_0) = \left[\begin{array}{ccc}
     H_{i,11}& H_{i,12}& H_{i,13}  \\
      H_{i,21}&  H_{i,22} & H_{i,23}\\
       H_{i,31}& H_{i,32}&  H_{i,33}
\end{array}
\right]. 
\eeq
In equation (\ref{x0}), $z_{(1),0}$ is the desired state vector of the robot system, and $z_{(2)0} $ is the desired state vector of the Koopman system when the desired $z_{(1)0} $ is achieved by $z_{(1)}$.
Since $U-U_0 = U$ is expected to converge to $0_{1\times 6}$, we may ignore the term $U^TH_{i,33}U$ in $(x-x_0)^T H_{f_{i}}(x_0)(x-x_0)$.

The expansion of  $(x-x_0)^T H_{f_{i}}(x_0)(x-x_0)$ can be rewritten as 
\begin{equation}
    \begin{split}
    &(x-x_0)^T H_{f_{i}}(x_0)(x-x_0) \\=& (z_{(1)}(k)-z_{(1)0})^T H_{i,11} (z_{(1)}(k)-z_{(1)0}) + (z_{(2)}(k)-z_{(2)0})^T H_{i,21} (z_{(1)}(k)-z_{(1)0})\\
    & +U^T(k) H_{i,31} (z_{(1)}(k)-z_{(1)0}) + (z_{(1)}(k)-z_{(1)0})^T H_{i,12} (z_{(2)}(k)-z_{(2)0})\\
    & + (z_{(2)}(k)-z_{(2)0})^T H_{i,22} (z_{(2)}(k)-z_{(2)0}) + U^T(k) H_{i,32} (z_{(2)}(k)-z_{(2)0})\\
    & + (z_{(1)}(k)-z_{(1)0})^T H_{i,13} U(k) + (z_{(2)}(k)-z_{(2)0})^T H_{i,23} U(k)+ \cancel{U^T(k) H_{i,33} U(k)}.
    \end{split}
\end{equation}

Since $H_i = H_i^T$, that is, $H_{i,21}^T = H_{i,12}$, $H_{i,31}^T = H_{i,13}$ and $H_{i,32}^T = H_{i,23}$, we have
\beq
U^T(k) H_{i,31} (z_{(1)}(k)-z_{(1)0}) =  (z_{(1)}(k)-z_{(1)0})^T H^T_{i,31} U(k) = (z_{(1)}(k)-z_{(1)0})^T H_{i,13} U(k),
\eeq
\beq
U^T(k) H_{i,32} (z_{(2)}(k)-z_{(2)0}) =  (z_{(2)}(k)-z_{(2)0})^T H^T_{i,32} U(k) = (z_{(2)}(k)-z_{(2)0})^T H_{i,23} U(k),
\eeq
\beq
(z_{(2)}(k)-z_{(2)0})^T H_{i,21} (z_{(1)}(k)-z_{(1)0})= (z_{(1)}(k)-z_{(1)0})^T H_{i,12} (z_{(2)}(k)-z_{(2)0}).
\eeq

For the parametrization, we find some colmumn parameter vectors $\phi_{ij},\;j=1,2,3,4,5,$ and column signal vectors $g_{i1}(\cdot)$, $g_{i2}(\cdot,\cdot)$, $g_{i3}(\cdot)$, $g_{i4}(\cdot,\cdot)$, $g_{i5}(\cdot,\cdot)$, express
\beq
(z_{(1)}(k)-z_{(1)0})^T H_{i,11} (z_{(1)}(k)-z_{(1)0}) = \phi_{i1}^Tg_{i1}(z_{(1)}(k)-z_{(1)0}),
\eeq
\beq
(z_{(2)}(k)-z_{(2)0})^T H_{i,21} (z_{(1)}(k)-z_{(1)0}) = \phi_{i2}^Tg_{i2}(z_{(2)}(k)-z_{(2)0},z_{(1)}(k)-z_{(1)0}),
\eeq
\beq
(z_{(2)}(k)-z_{(2)0})^T H_{i,22} (z_{(2)}(k)-z_{(2)0}) = \phi_{i3}^Tg_{i3}(z_{(2)}(k)-z_{(2)0}),
\eeq
\beq
(z_{(1)}(k)-z_{(1)0})^T H_{i,13} U(k) = \phi_{i4}^Tg_{i4}(z_{(1)}(k)-z_{(1)0},U(k)),
\eeq
\beq
(z_{(2)}(k)-z_{(2)0})^T H_{i,23} U(k) = \phi_{i5}^Tg_{i5}(z_{(2)}(k)-z_{(2)0},U(k)).
\eeq
Then, the term $(x-x_0)^T H_{f_{i}}(x_0)(x-x_0)$ can be expressed as 
\beq
\begin{split}
    &(x-x_0)^T H_{f_{i}}(x_0)(x-x_0)\; \text{without} \; U^TH_{i,33}U \\= & \phi_{i1}^Tg_{i1}(z_{(1)}(k)-z_{(1)0}) + 2\phi_{i2}^Tg_{i2}(z_{(2)}(k)-z_{(2)0},z_{(1)}(k)-z_{(1)0})
    +\phi_{i3}^Tg_{i3}(z_{(2)}(k)-z_{(2)0})  \\
    &+2\phi_{i4}^Tg_{i4}(z_{(1)}(k)-z_{(1)0},U(k))+ 2\phi_{i5}^Tg_{i5}(z_{(2)}(k)-z_{(2)0},U(k)). 
\end{split}
\eeq
For the $3$-robot system, we have the following expressions\footnote{$(V)_{l}$ denotes the $l$th component of vector $V$. For example $(z_{(1)}(k)-z_{(1)0})_{2}$ denotes the second component of the vector $z_{(1)}(k)-z_{(1)0}$.}:
\beq
g_{i1}(z_{(1)}(k)-z_{(1)0}) = \left [ \begin{array}{c}
     (z_{(1)}(k)-z_{(1)0})_{1} \cdot (z_{(1)}(k)-z_{(1)0}) \\
     (z_{(1)}(k)-z_{(1)0})_{2} \cdot (z_{(1)}(k)-z_{(1)0})\\
     \vdots\\ 
     (z_{(1)}(k)-z_{(1)0})_{12} \cdot (z_{(1)}(k)-z_{(1)0})
\end{array}       
\right ]\in R^{144},
\label{gis}
\eeq
\beq
g_{i2}(z_{(2)}(k)-z_{(2)0},z_{(1)}(k)-z_{(1)0}) = \left [ \begin{array}{c}
     (z_{(1)}(k)-z_{(1)0})_{1} \cdot (z_{(2)}(k)-z_{(2)0}) \\
     (z_{(1)}(k)-z_{(1)0})_{2} \cdot (z_{(2)}(k)-z_{(2)0}) \\
     \vdots\\ 
     (z_{(1)}(k)-z_{(1)0})_{12} \cdot (z_{(2)}(k)-z_{(2)0}) 
\end{array}       
\right ]\in R^{216},
\eeq
\beq
g_{i3}(z_{(2)}(k)-z_{(2)0}) = \left [ \begin{array}{c}
     (z_{(2)}(k)-z_{(2)0})_{1} \cdot (z_{(2)}(k)-z_{(2)0}) \\
      (z_{(2)}(k)-z_{(2)0})_{2} \cdot (z_{(2)}(k)-z_{(2)0}) \\
     \vdots\\ 
      (z_{(2)}(k)-z_{(2)0})_{18} \cdot (z_{(2)}(k)-z_{(2)0}) 
\end{array}       
\right ]\in R^{324},
\eeq
\beq
g_{i4}(z_{(1)}(k)-z_{(1)0},U(k)) = \left [ \begin{array}{c}
     (U(k))_{1} \cdot (z_{(1)}(k)-z_{(1)0}) \\
      (U(k))_{2} \cdot (z_{(1)}(k)-z_{(1)0}) \\
     \vdots\\ 
     (U(k))_{6} \cdot (z_{(1)}(k)-z_{(1)0})
\end{array}       
\right ]\in R^{72},
\eeq
\beq
g_{i5}(z_{(2)}(k)-z_{(2)0},U(k)) = \left [ \begin{array}{c}
     (U(k))_{1} \cdot (z_{(2)}(k)-z_{(2)0}) \\
      (U(k))_{2} \cdot (z_{(2)}(k)-z_{(2)0}) \\
     \vdots\\ 
     U(k)_{6} \cdot (z_{(2)}(k)-z_{(2)0})
\end{array}       
\right ]\in R^{108}.
\label{gie}
\eeq

Then, for (\ref{gis})-(\ref{gie}), we have 
\beq
\begin{split}
    \phi_{i1} \in R^{144},
    \phi_{i2} \in R^{216},
    \phi_{i3} \in R^{324},
    \phi_{i4} \in R^{72},
    \phi_{i5} \in R^{108},\;i\in \{1,2,3,\cdots, 18 \}.\\
\end{split}
\eeq
It can be verified that for all $i$, we have
\begin{equation}
    g_{ij}(\cdot) = g_j(\cdot), j=1,3,\;g_{ij}(\cdot,\cdot) = g_j(\cdot,\cdot), j=2,4,5. \; 
\end{equation}

Finally, we obtain the model for the nonlinear utility function components
\begin{equation}
\begin{split}
        z_{(2)}(k+1) =& A_{21}(z_{(1)}(k)-z_{(1)0}) + A_{22}(z_{(2)}(k)-z_{(2)0}) + B_2U(k)\\
&+f(z_{(1)0},\,z_{(2)0}) + \Phi_{1}^Tg_{1}(z_{(1)}(k)-z_{(1)0}) + \Phi_{2}^Tg_{2}(z_{(2)}(k)-z_{(2)0},z_{(1)}(k)-z_{(1)0})\\
   & +\Phi_{3}^Tg_{3}(z_{(2)}(k)-z_{(2)0})  
    +\Phi_{4}^Tg_{4}(z_{(1)}(k)-z_{(1)0},U(k))\\&+ \Phi_{5}^Tg_{5}(z_{(2)}(k)-z_{(2)0},U(k)) + \eta(k)
\end{split}
\label{bilinear_model}
\end{equation}
for some error vector $\eta(k) \in R^{18}$, where the rows of $\Phi_1^T$ are $0.5\phi_{i1}^T$, the rows of $\Phi_2^T$ are $\phi_{i2}^T$, the rows of $\Phi_3^T$ are $0.5\phi_{i3}^T$, the rows of $\Phi_4^T$ are $\phi_{i4}^T$, and the rows of $\Phi_5^T$ are $\phi_{i5}^T$, for $i=1,2,\cdots,18 $ (for the $3$-robot system). Thus, we have that $\Phi_1\in R^{144\times 18}$, $\Phi_2\in R^{216\times 18}$, $\Phi_3\in R^{324\times 18}$, $\Phi_4\in R^{72\times 18}$ and $\Phi_5\in R^{108 \times 18}$.  Correspondingly, the unknown parameters $A_{21}$, $A_{22}$, $B_2$, $f_0 = f(z_{(1)0},z_{(2)0})$, $\Phi_1^T$, $\Phi_2^T$, $\Phi_3^T$, $\Phi_4^T$ and $\Phi_5^T$ are estimated as $\hat{A}_{21}$, $\hat{A}_{22}$, $\hat{B}_2$, $\hat{f}_0 $, $\hat{\Phi}_1^T$, $\hat{\Phi}_2^T$, $\hat{\Phi}_3^T$, $\hat{\Phi}_4^T$ and $\hat{\Phi}_5^T$ by the following iterative algorithm. 

\bigskip
{\bf Parametrization}. To develop an adaptive algorithm for
generating the  iterative (online) estimates of $A_{21}$, $A_{22}$, $B_2$, $f_0 = f(z_{(1)0},z_{(2)0})$, $\Phi_1^T$, $\Phi_2^T$, $\Phi_3^T$, $\Phi_4^T$ and $\Phi_5^T$, we define 
\beq
y(k) = z_{(2)}(k+1) \in R^{18}
\label{y(k)(z_2)_bilinear_raw}
\eeq
\begin{equation}
    \zeta(k) = [(z_{(1)}(k)-z_{(1)0})^T,(z_{(2)}(k)-z_{(2)0})^T,U^T(k), 1, g_1^T(\cdot),g_2^T(\cdot,\cdot),g_3^T(\cdot),g_4^T(\cdot,\cdot),g_5^T(\cdot,\cdot)]^T \in R^{901} .
    \label{zeta_bilinear}
\end{equation}
\begin{equation}
    {\Theta} (k)= [{A}_{21}(k),{A}_{22}(k),{B}_{2}(k), {f}(z_{(1)0},\,z_{(2)0})(k),{\Phi}_1^T(k),{\Phi}_2^T(k),{\Phi}_3^T(k),{\Phi}_4^T(k),{\Phi}_5^T(k)]^T\in R^{ 901 \times 18}. 
\end{equation}
and rewrite (\ref{y(k)(z_2)_bilinear_raw}) as
\beq
y(k) = \Theta^T \zeta(k) + \eta_{(2)}(k),\;
\Theta^T = \left[\begin{array}{c}
\theta_1^T\\
\vdots\\
\theta_{18}^T
\end{array}
\right].
\label{y(k)_bilinear_paraed}
\eeq
With the estimates \begin{equation}
    \hat{\Theta} (k)= [\hat{A}_{21}(k),\hat{A}_{22}(k),\hat{B}_{2}(k), \hat{f}(z_{(1)0},\,z_{(2)0})(k),\hat{\Phi}_1^T(k),\hat{\Phi}_2^T(k),\hat{\Phi}_3^T(k),\hat{\Phi}_4^T(k),\hat{\Phi}_5^T(k)]^T\in R^{901\times 18} 
    \label{theta_hat_bilinear}
\end{equation} of the unknown $\Theta$, the approximated linear model for the nonlinear utility function (\ref{nonlinear_zk+1}) is 
\beq
 \begin{split}
        \hat{z}_{(2)}(k+1) = & \hat{A}_{21}(k)(z_{(1)}(k)-z_{(1)0}) + \hat{A}_{22}(k)(z_{(2)}(k)-z_{(2)0}) + \hat{B}_2(k)U(k)\\
&+\hat{f}(z_{(1)0},\,z_{(2)0})(k) + \hat{\Phi}_{1}^T(k)g_{1}(z_{(1)}(k)-z_{(1)0})\\& + \hat{\Phi}_{2}^T(k)g_{2}(z_{(2)}(k)-z_{(2)0},z_{(1)}(k)-z_{(1)0})
    +\hat{\Phi}_{3}^T(k)g_{3}(z_{(2)}(k)-z_{(2)0})  \\
    &+\hat{\Phi}_{4}^T(k)g_{4}(z_{(1)}(k)-z_{(1)0},U(k))+ \hat{\Phi}_{5}^T(k)g_{5}(z_{(2)}(k)-z_{(2)0},U(k)).
    \end{split}
\label{z(2)(k+1)appr_bilinear}
\eeq

\subsection{Optimal Control Problems}
In this report, the control objective is to ensure the robots reach their respective target positions without collisions between robots and robots or between robots and the circular wall. In this subsection, the specific control design solutions for the control objective are discussed after the introduction of the nonlinear programming control problem. 

\bigskip
{\bf Nonlinear programming control problem}. The game theory based method proposed in \cite{wetal20} can achieve the control objective by maximizing the nonlinear utility function (\ref{utility2}), \begin{equation*}
    \begin{split}
    u(k) &= \sum_{i=1}^3u_{i}(k),\\
\end{split}
\end{equation*}
where $u_{i}(k) = \sum_{j=1}^6 \omega_i^{(j)} \phi_{i}^{(j)}(X(k)), i = 1,2,3$. As introduced before, the nonlinear utility function components $\phi_{i}^{(j)}(X(k))$ are
computed by equations (\ref{phi_c1})-(\ref{phi_c6}) according to the robot state vector $X(k)$ depending on $s_{l}(k) = \{r_{l}(k), v_{l}(k)\} \; (l = 1,2,3)$. 

Although the nonlinear utility function based on control design can achieve the control objective, the nonlinear programming needs quite a long time to find the optimal solution, which indicates that the control design cannot be used in practical situations. 

\subsubsection{Linear Programming Control Problem Formulations}
In this report, the linear approximation model and the bilinear approximation model for the nonlinear utility function components are proposed to speed up finding the optimal control input for the system to achieve the control objective. With the approximated models, the method to find the optimal control input, maximizing the utility function, is transformed into a linear programming problem. The solutions based on the linear model approximation and bilinear model approximation are presented in the following. 

\bigskip
{\bf Solution with linear model approximation}. Based on the linear model approximation, $\hat{z}_{(2)}(k)$, for the linear Koopman system model of the utility function components in the $3$-robot system $z_{(2)}(k+1) = A_{21} z_{(1)}(k) + A_{22} z_{(2)}(k) + B_2  U(k) +
\eta_{(2)}(k)\in R^{18}$, the utility function for the whole $3$-robot system can be expressed as 
\begin{equation}
\begin{split}
    u(k) &= w^T\hat{z}_{(2)}(k+1)\\
    &= w^T(\hat{A}_{21} {z}_{(1)}(k) + \hat{A}_{22} {z}_{(2)}(k) + \hat{B}_2  U(k)),
\end{split}
\end{equation}
where $w = [w_1,w_2 ,w_3]^T$, with $w_i=[w_i^{(1)},w_i^{(2)},w_i^{(3)},w_i^{(4)},w_i^{(5)},w_i^{(6)}]^T, i = 1,2,3$, is a chosen constant vector, and $\hat{z}_{(2)}(k+1)$ is the estimates from the linear model identification algorithm.  To maximize the linear expression of the utility function, the solution for the control input, based on the approximate linear model for the utility function components, is denoted as

\begin{equation}
    U^*(k) = \arg \max _{U(k)\in U_c} w^T\hat{z}_{(2)}(k+1),
    \label{sol_inputfromlinear}
\end{equation}
where $w$ is a chosen constant vector and the $U_c = \{U(k)|U_i(k)\in(-4,3),i = 1,\cdots, 6\}$ is the upper bound and the lower bound of the control input.

\bigskip
{\bf Solution with bilinear model approximation}. Similarly, with the estimation, $\hat{z}_{(2)}(k+1)$, of the bilinear Koopman system model (\ref{z(2)(k+1)appr_bilinear}), 
the utility function for the whole $3$-robot system can be expressed as 
\begin{equation}
    u(k) = w^T\hat{z}_{(2)}(k+1),
\end{equation}
where $w = [w_1,w_2, w_3]^T$, with $w_i=[w_i^{(1)},w_i^{(2)},w_i^{(3)},w_i^{(4)},w_i^{(5)},w_i^{(6)}]^T, i = 1,2,3$, is a chosen constant vector and $\hat{z}_{(2)}(k+1)$ is computed from (\ref{z(2)(k+1)appr_bilinear}). To maximize the linear expression of the utility function, the solution for the control input, based on the approximate linear model for the utility function components, is denoted as

\begin{equation}
    U^*(k) = \arg \max _{U(k)\in U_c} w^T\hat{z}_{(2)}(k+1),
    \label{sol_inputfrombilinear}
\end{equation}
where $w$ is a chosen constant vector and the $U_c = \{U(k)|U_i(k)\in(-4,3),i = 1,\cdots 6\}$ is the upper bound and the lower bound of the control input. 

\subsubsection{Matlab Implementations}
In Matlab, there is a built function $\mathtt{linprog}$ that can solve linear programming problems. The specific procedures for the linear model based optimal control design are listed as an example
\footnote{ \url{https://www.mathworks.com/help/releases/R2020b/optim/ug/linprog.html?lang=en}}:
\begin{enumerate}
    \item[\textbf{Step 1}:] Create the 6-by-1 optimization variable vector ensuring every element belongs to $[-4,3]$ by the following Matlab codes
    \begin{equation}
        \mathtt{x = optimvar('x',6,'LowerBound',-4,'UpperBound',3)};
        \label{LP_start}
    \end{equation}
    \item[\textbf{Step 2}:] Create the optimization problem and identify the optimization objective, which is maximizing the linear model based utility function $w^T(\hat{A}_{21}(k)z_{(1)}(k)+\hat{A}_{22}(k)z_{(2)}(k)+\hat{B}_2(k)U(k))$, by codes
    \begin{equation}
    \begin{split}
                \mathtt{prob = optimproblem(}& \mathtt{'Objective',w' * (A_{21}(k) * z_{(1)}(k) + A_{22}(k) * z_{(2)}(k) + B_2(k) * x),}\\
                &\mathtt{'ObjectiveSense','max')};
    \end{split}
    \end{equation}
    
    \item[\textbf{Step 3}:] Find the solution by codes
    \begin{equation}
        \begin{split}
           & \mathtt{problem}\, \mathtt{= prob2struct(prob)},\\
            &\mathtt{U} \, \mathtt{= linprog(problem)},
        \end{split}
        \label{LP_end}
    \end{equation}
    Vector $u$ is the solution for the created linear programming problem.
\end{enumerate}

\setcounter{equation}{0}
\section{Adaptive Parameter Estimation}
\label{Sec_ParaEst}
This section introduces the adaptive parameter algorithm to estimate the unknown parameters of the linear approximation model or the blinear approximation model based on the least-square. 
\subsection{Least-Squares Formulation}
According to (\ref{y(k)_linear_paraed}) and (\ref{y(k)_bilinear_paraed}), the objective is to find the vector $\hat{\Theta}(k)$ to minimize the error vector $\hat{\Theta}^{T}(k) \zeta(\tau) -
y(\tau)$. In \cite{tzh22a}, $J_3$ (for an iterative
(online) algorithm) 
\bea
J_3 = J_3(\hat{\Theta}) \ts = \ts \frac{1}{2} \sum_{\tau = k_0}^{k-1} 
\frac{1}{\rho} (\hat{\Theta}^{T}(k) \zeta(\tau) - y(\tau))^{T}
(\hat{\Theta}^{T}(k) \zeta(\tau) - y(\tau))\nn \\
\ts \ts +\frac{1}{2}\tr[(\hat{\Theta}(k) - \Theta_{0})^{T} P_{0}^{-1} (\hat{\Theta}(k) -
\Theta_{0})],\;P_{0} = P_{0}^{T} \left(\in R^{36 \times 36}\right) >
0,\;\rho > 0,
\label{J_3}
\eea
is proposed from $J_2$ (for a batch-data algorithm)
\beq
J_2 = \sum_{\tau=k_0}^{k-1} (\hat{\Theta}^{T}(k) \zeta(\tau) -
y(\tau))^{T} (\hat{\Theta}^{T}(k) \zeta(\tau) - y(\tau)),
\eeq
to be our cost function for the iterative parameter estimation algorithm. 
Such an adaptive algorithm is derived based on a modified version $J_3$ of the cost function for generating the online estimate
$\hat{\Theta}(k)$, using the real-time updated measurement data $y(k)$ and $\zeta(k)$. 

In this cost function $J_3$, $\Theta_{0}$ is the initial estimate of
$\Theta$: $\hat{\Theta}(0) = \Theta_0$, $P_{0} = P_{0}^{T} > p_0
\,I$ is a parameter matrix for some chosen scalar $p_0 > 0$ (which is
typically a large number) (the term $\frac{1}{2}\tr[(\hat{\Theta}(k) -
\Theta_{0})^{T} P_{0}^{-1} (\hat{\Theta}(k)-\Theta_{0})]$ represents
a penalty on the initial estimate $\hat{\Theta}_0$), and $\rho > 0$ is
a design parameter (which is typically a small number).

To find the vector $\hat{\Theta}(k)$ minimizing the cost function $J_3$, a normalized least-squares algorithm is proposed in the next subsection. More details about the least-squares algorithm are in \cite{t03}.

\subsection{Adaptive Parameter Estimation Algorithm}
Minimization of the above $J_3$ leads to the following iterative
algorithm to generate the online estimate $\hat{\Theta}(k)$, as the
new measurements $y(k)$ and $\zeta(k)$ are collected, for $k = 1,
\ldots$. This subsection presents the adaptive parameter algorithm and its applications to the linear and the bilinear model. 

\subsubsection{Adaptive Algorithm}
To find the desired parameter vector $\hat{\Theta}(k)$, we propose an adaptive algorithm. At first, we define the difference between the linear model outputs and the original nonlinear utility function components to be the estimation error
\begin{equation}
    \epsilon(k) = \hat{\Theta}^T(k) \zeta(k)-y(k),
\label{epsilon(k)}
\end{equation}
at first, where $y(k)=z_{(2)}(k+1)\in R^{18}$. Then, the estimate $\hat{\Theta}(k)$ of the parameter matrix is generated from the adaptive law
\begin{equation}
    \hat{\Theta}(k+1) = \hat{\Theta}(k)-\frac{P(k-1) \zeta(k)
  \epsilon^T(k)}{m^2(k)},\;k = 0, 1, 2,\ldots,
\label{Thetak+1}
\end{equation}
with $\hat{\Theta}(0)$ chosen, where
\beq
m(k) = \sqrt{\rho + \zeta^{T}(k)P(k-1)\zeta(k)},\;\rho > 0,
\eeq
\beq
P(k) = P(k-1)-\frac{P(k-1) \zeta(k) \zeta^{T}(k)
P(k-1)}{m^{2}(k)},\;k = 0, 1, 2, \ldots,
\label{Pk1}
\eeq
with $P(-1) = P_{0} = P^{T}_{0} > 0$ chosen.

\bigskip
\par \textbf{Matrix $P(k)$ resetting.} To improve the estimation results, we reset the matrix $P(k)=P_0$ every fixed time interval. 

\bigskip
\textbf{Remark}. The sizes of $\Theta(k)$ and $P(k)$ depend on the sizes of $\zeta(k)$ in different model.
\subsubsection{Application to Linear Model}
For the unknown parameters $\Theta^T = [A_{21}, A_{22}, B_2] \in R^{18 \times 36}$ in the linear model
\begin{equation*}
    \begin{split}
        z_{(2)}(k+1) &= A_{21} z_{(1)}(k) + A_{22} z_{(2)}(k) + B_2  U(k) +
\eta_{(2)}(k)\\
&= \Theta^T\zeta(k)+\eta_{(2)}(k),
    \end{split}
\end{equation*}
where 
\begin{equation}
\zeta(k) = [z_{(1)}^T(k), z_{(2)}^T(k), U^T(k)]^T \in R^{36},    
\end{equation}
we have the estimate of those parameters \begin{equation}
    \hat{\Theta}^T(k) = [\hat{A}_{21}(k), \hat{A}_{22}(k), \hat{B}_2(k)] \in R^{18 \times 36}.
\end{equation}
Then, the adaptive algorithm for the linear approximation model has the form (\ref{epsilon(k)})-(\ref{Pk1}) with $\zeta(k)\in R^{36}$, $\hat{\Theta}^T(k)\in R^{18\times36 }$ and $P\in R^{36\times36}$.

\subsubsection{Application to Bilinear Model}
For the unknown parameters $
    {\Theta}= [{A}_{21},{A}_{22},{B}_{2}, {f}(z_{(1)0},\,z_{(2)0}),{\Phi}_1^T,{\Phi}_2^T,{\Phi}_3^T,{\Phi}_4^T,{\Phi}_5^T]^T\in R^{ 901\times 18 }$ in the bilinear model
    \begin{equation*}
        \begin{split}
        z_{(2)}(k+1) =& A_{21}(z_{(1)}(k)-z_{(1)0}) + A_{22}(z_{(2)}(k)-z_{(2)0}) + B_2U(k)\\
&+f(z_{(1)0},\,z_{(2)0}) + \Phi_{1}^Tg_{1}(z_{(1)}(k)-z_{(1)0}) + \Phi_{2}^Tg_{2}(z_{(2)}(k)-z_{(2)0},z_{(1)}(k)-z_{(1)0})\\&
    +\Phi_{3}^Tg_{3}(z_{(2)}(k)-z_{(2)0})  
    +\Phi_{4}^Tg_{4}(z_{(1)}(k)-z_{(1)0},U(k))\\&+ \Phi_{5}^Tg_{5}(z_{(2)}(k)-z_{(2)0},U(k)) + \eta_{(2)}(k)\\
    =&\Theta^T\zeta(k)+\eta_{(2)}(k),
\end{split}
    \end{equation*}
    where
    \begin{equation}
    \zeta(k) = [(z_{(1)}(k)-z_{(1)0})^T,(z_{(2)}(k)-z_{(2)0})^T,U^T(k), 1, g_1^T(\cdot),g_2^T(\cdot,\cdot),g_3^T(\cdot),g_4^T(\cdot,\cdot),g_5^T(\cdot,\cdot)]^T \in R^{901} ,
    \end{equation}
    we have the estimate of those parameters 
    \begin{equation}
    \hat{\Theta} (k)= [\hat{A}_{21}(k),\hat{A}_{22}(k),\hat{B}_{2}(k), \hat{f}(z_{(1)0},\,z_{(2)0})(k),\hat{\Phi}_1^T(k),\hat{\Phi}_2^T(k),\hat{\Phi}_3^T(k),\hat{\Phi}_4^T(k),\hat{\Phi}_5^T(k)]^T\in R^{ 901\times 18 }    .
    \end{equation}
     
    The iterative algorithm for the parameter estimate matrix $\hat{\Theta}(k)\in R^{901\times 18 }$ to minimize the estimation error $\epsilon(k)$ has the form (\ref{epsilon(k)})-(\ref{Pk1}) with $\zeta(k)\in R^{901}$, $\hat{\Theta}(k)\in R^{901\times18  }$ and $P\in R^{901\times 901 }$.

Then, the approximated linear model for the nonlinear utility function components (\ref{nonlinear_zk+1}) can be expressed as (\ref{z(2)(k+1)appr_bilinear}).

\section{Decentralized Model Approximation and Control}
\label{Sec_DecThe}
Since centralized control strategies cannot solve the state explosion problem of the system containing too many agents, decentralized control strategies are more proper for these cases. This section formulates a decentralized bilinear model approximation and an optimal control design based on the approximated bilinear model. The parameters in the approximation models will update separately for the three robots. At the same time, every robot shares its own position and speed information with all other robots. With the approximation models for every robot, the optimal inputs for each robot will be computed separately by maximizing its individual estimated utility function. 
\subsection{Decentralized Bilinear Koopman System Model }
This subsubsection presents the detailed decentralized Koopman system models to acquire a bilinear representation for the multi-robot system with nonlinear utility functions. 

\bigskip
\textbf{{Decentralized Koopman System Variables}}. For the decentralized Koopman system of every single robot $i$, the variables $z_{(1),i}(k)$ and $z_{(2),i}(k)$ are parts of the system variables in the centralized Koopman system introduced \eqref{z1_def} and \eqref{z2_def},
\beq
z_{(1),i}(k) = [r_i^T(k),v_i^T(k)]^T\in R^{4}, \; i = 1,2,3,
\label{z1_def_d}
\eeq
\beq
z_{(2),i}(k) = [\phi_{i}^{(1)}(X(k)),\phi_{i}^{(2)}(X(k)),\phi_{i}^{(3)}(X(k)),\phi_{i}^{(4)}(X(k)),\phi_{i}^{(5)}(X(k)),\phi_{i}^{(6)}(X(k))]^T\in R^{6},
\label{z2_def_d}
\eeq
where $r_s(k)$ and $v_s(k)$ are the position and the velocity information, defined in \eqref{3r_robots_indiv}, for Robot $s$, $\phi_{s}^{(i)}(X(k))$ denotes the $i$th utility function component, defined in \eqref{phi_c1}-\eqref{phi_c6}, of Robot $i$, and $X(k)$, defined in \eqref{3r_z1}, is the vector containing the position and velocity information of all three robots. 

Correspondingly, the input $U_i(k)$ of the decentralized single robot system is also defined in \eqref{3r_robots_indiv}.

\bigskip
\textbf{Decentralized Bilinear Model for Nonlinear Utility Functions}. The decentralized bilinear model is similar to the centralized bilinear model \eqref{bilinear_model}
\begin{equation}
\begin{split}
        z_{(2),i}(k+1) =& A_{21}(z_{(1),i}(k)-z_{(1)0,i}) + A_{22}(z_{(2),i}(k)-z_{(2)0,i}) + B_2U_i(k)\\
&+f(z_{(1)0,i},\,z_{(2)0,i}) + \Phi_{1}^Tg_{1}(z_{(1),i}(k)-z_{(1)0,i}) + \Phi_{2}^Tg_{2}(z_{(2),i}(k)-z_{(2)0,i},z_{(1),i}(k)-z_{(1)0,i})\\
   & +\Phi_{3}^Tg_{3}(z_{(2),i}(k)-z_{(2)0,i})  
    +\Phi_{4}^Tg_{4}(z_{(1),s}(k)-z_{(1)0,i},U_i(k))\\&+ \Phi_{5}^Tg_{5}(z_{(2),i}(k)-z_{(2)0,i},U_i(k)) + \eta(k),
\end{split}
\label{bilinear_model_d}
\end{equation}
for some error vector $\eta(k) \in R^{6}$, where $z_{(1)0,i}$ is the desired state vector of the robot system, $z_{(2)0,i} $ is the desired state vector of the Koopman system when the desired $z_{(1)0,i} $ is achieved by $z_{(1),i}$,
the vectors $g_i, \;i\in1,2,3,4,5$ are defined as 
\beq
g_{1}(z_{(1),s}(k)-z_{(1)0,i}) = \left [ \begin{array}{c}
     (z_{(1),s}(k)-z_{(1)0,i})_{1} \cdot (z_{(1),s}(k)-z_{(1)0,i}) \\
     (z_{(1),s}(k)-z_{(1)0,i})_{2} \cdot (z_{(1),s}(k)-z_{(1)0,i})\\
     \vdots\\ 
     (z_{(1),s}(k)-z_{(1)0,i})_{4} \cdot (z_{(1),s}(k)-z_{(1)0,i})
\end{array}       
\right ]\in R^{16},
\eeq

\beq
g_{2}(z_{(2),i}(k)-z_{(2)0,i},z_{(1),s}(k)-z_{(1)0,i}) = \left [ \begin{array}{c}
     (z_{(1),s}(k)-z_{(1)0,i})_{1} \cdot (z_{(2),i}(k)-z_{(2)0,i}) \\
     (z_{(1),s}(k)-z_{(1)0,i})_{2} \cdot (z_{(2),i}(k)-z_{(2)0,i}) \\
     \vdots\\ 
     (z_{(1),s}(k)-z_{(1)0,i})_{4} \cdot (z_{(2),i}(k)-z_{(2)0,i}) 
\end{array}       
\right ]\in R^{24},
\eeq
\beq
g_{3}(z_{(2),i}(k)-z_{(2)0,i}) = \left [ \begin{array}{c}
     (z_{(2),i}(k)-z_{(2)0,i})_{1} \cdot (z_{(2),i}(k)-z_{(2)0,i}) \\
      (z_{(2),i}(k)-z_{(2)0,i})_{2} \cdot (z_{(2),i}(k)-z_{(2)0,i}) \\
     \vdots\\ 
      (z_{(2),i}(k)-z_{(2)0,i})_{6} \cdot (z_{(2),i}(k)-z_{(2)0,i}) 
\end{array}       
\right ]\in R^{36},
\eeq
\beq
g_{4}(z_{(1),s}(k)-z_{(1)0,i},U_i(k)) = \left [ \begin{array}{c}
     (U_i(k))_{1} \cdot (z_{(1),s}(k)-z_{(1)0,i}) \\
      (U_i(k))_{2} \cdot (z_{(1),s}(k)-z_{(1)0,i}) 
\end{array}       
\right ]\in R^{8},
\eeq
\beq
g_5(z_{(2),i}(k)-z_{(2)0,i},U_i(k)) = \left [ \begin{array}{c}
     (U_i(k))_{1} \cdot (z_{(2),i}(k)-z_{(2)0,i}) \\
      (U_i(k))_{2} \cdot (z_{(2),i}(k)-z_{(2)0,i}) 
\end{array}       
\right ]\in R^{12},
\eeq
and $A_{21}\in R^{6 \times 4}$, $A_{22}\in R^{6 \times 6}$, $B_2\in R^{6 \times 2}$, $\Phi_1\in R^{16\times 6}$, $\Phi_2\in R^{24\times 6}$, $\Phi_3\in R^{36\times 6}$, $\Phi_4\in R^{8\times 6}$ and $\Phi_5\in R^{12 \times 6}$ are the unknown parameters which are estimated as $\hat{A}_{21}$, $\hat{A}_{22}$, $\hat{B}_2$, $\hat{f}_0 $, $\hat{\Phi}_1^T$, $\hat{\Phi}_2^T$, $\hat{\Phi}_3^T$, $\hat{\Phi}_4^T$ and $\hat{\Phi}_5^T$ by the following iterative algorithm.

\subsection{Adaptive Parameter Estimation for the Decentralized Model }
This subsection introduces the model parametrization and the adaptive parameter estimation algorithm application for the decentralized bilinear model. 

\bigskip
{\bf Parametrization}. To develop an adaptive algorithm for
generating the  iterative (online) estimates of $A_{21}$, $A_{22}$, $B_2$, $f_0 = f(z_{(1)0,i},z_{(2)0,i})$, $\Phi_1^T$, $\Phi_2^T$, $\Phi_3^T$, $\Phi_4^T$ and $\Phi_5^T$, we define 
\beq
y_i(k) = z_{(2),i}(k+1) \in R^{6}
\label{y(k)(z_2)_bilinear_raw_d}
\eeq
\begin{equation}
    \zeta_i(k) = [(z_{(1),s}(k)-z_{(1)0,i})^T,(z_{(2),i}(k)-z_{(2)0,i})^T,U_i^T(k), 1, g_1^T(\cdot),g_2^T(\cdot,\cdot),g_3^T(\cdot),g_4^T(\cdot,\cdot),g_5^T(\cdot,\cdot)]^T \in R^{109} .
    \label{zeta_bilinear_d}
\end{equation}
\begin{equation}
    {\Theta}_i (k)= [{A}_{21}(k),{A}_{22}(k),{B}_{2}(k), {f}(z_{(1)0,i},\,z_{(2)0,i})(k),{\Phi}_1^T(k),{\Phi}_2^T(k),{\Phi}_3^T(k),{\Phi}_4^T(k),{\Phi}_5^T(k)]^T\in R^{ 109 \times 6}. 
\end{equation}
and rewrite (\ref{y(k)(z_2)_bilinear_raw_d}) as
\beq
y_i(k) = \Theta^T_i\zeta_i(k) + \eta_{(2)}(k),\;
\Theta^T_i= \left[\begin{array}{c}
\theta_1^T\\
\vdots\\
\theta_{6}^T
\end{array}
\right].
\eeq
With the estimates \begin{equation}
    \hat{\Theta}_i (k)= [\hat{A}_{21}(k),\hat{A}_{22}(k),\hat{B}_{2}(k), \hat{f}(z_{(1)0,i},\,z_{(2)0,i})(k),\hat{\Phi}_1^T(k),\hat{\Phi}_2^T(k),\hat{\Phi}_3^T(k),\hat{\Phi}_4^T(k),\hat{\Phi}_5^T(k)]^T\in R^{109\times 6} 
    \label{theta_hat_bilinear_d}
\end{equation} of the unknown $\Theta$, the approximated bilinear model for the nonlinear utility function (\ref{nonlinear_zk+1}) is 
\beq
 \begin{split}
 \hat{z}_{(2),i}(k+1) = & \hat{\Theta}_i^T (k)\zeta_i(k).
    \end{split}
\label{z(2)(k+1)appr_bilinear_d}
\eeq

\bigskip
\textbf{Adaptive parameter estimation algorithm}. In Section 3.2, a least square parameter estimation algorithm \eqref{Thetak+1}-\eqref{Pk1} proposed to find the parameter matrix $\hat{\Theta}_i(k)$ which can minimize the error vector $\epsilon(k) = \hat{\Theta}_i^{T}(k) \zeta_i(k) -
y_i(k)$, where $y_i(k)=z_{(2),i}(k+1)\in R^{6}$. For the decentralized bilinear model, the parameter matrices and the signals are replaced by \eqref{theta_hat_bilinear_d} and \eqref{zeta_bilinear_d}. 

\subsection{Optimal Control Based on the Decentralized Model}
The game theory based method proposed in \cite{wetal20} can achieve the control objective by maximizing the nonlinear utility function (\ref{utility2}),
 $u_{i}(k) = \sum_{j=1}^6 \omega_i^{(j)} \phi_{i}^{(j)}(X(k)), i = 1,2,3$. As introduced before, the nonlinear utility function components $\phi_{i}^{(j)}(X(k))$ are
computed by equations (\ref{phi_c1})-(\ref{phi_c6}) according to the robot state vector $X(k)$ depending on $s_{l}(k) = \{r_{l}(k), v_{l}(k)\} \; (l = 1,2,3)$. 

With the estimation for Robot $i$ from the bilinear Koopman system model (\ref{z(2)(k+1)appr_bilinear}), $\hat{z}_{(2),i}(k+1)$, 
the utility function for the single robot system can be expressed as 
\begin{equation}
    u_i(k) = w_i^T\hat{z}_{(2),i}(k+1),
\end{equation}
where $w_i $ is a chosen constant vector and $\hat{z}_{(2),i}(k+1)$ is computed from (\ref{z(2)(k+1)appr_bilinear}). To maximize the bilinear expression of the utility function, the solution for the control input, based on the approximate bilinear model for the utility function components, is denoted as

\begin{equation}
    U^*_i(k) = \arg \max _{U_i(k)\in U_{ci}} w_i^T\hat{z}_{(2),i}(k+1),
    \label{sol_inputfrombilinear_d}
\end{equation}
where $w_i$ is a chosen constant vector and the $U_{ci} = \{U_i(k)|(U_i(k))_l\in(-4,3), l = 1,2\}$ is the upper bound and the lower bound of the control input.

\setcounter{equation}{0}
\section{Simulation Studies}
\label{Sec_Simu}
This section presents the simulation study of the proposed system identification method and the control design. The 3-robot system is applied to examine the linear and bilinear model identification performance and the performance of the control designs based on the approximated models. Additionally, the decentralized bilinear model based control design is also simulated.

\subsection{Linear Model Identification of 3-Robot System}
\label{Sec_simuLinID}
In this subsection, the simulation system and the simulation results for the linear model identification of the 3-robot system are demonstrated sequentially. The estimation errors are examined to check the performance of the model (system) identification. The selected parameter components and the norms of the parameter matrices are analyzed for the parameter results.

\subsubsection{Simulation system}
In this part, the basic simulation construction and data collection logic are introduced. 

\bigskip
\textbf{System state variables}. System state variables $z(k)$ are constructed with natural state variables and utility function components. For the natural state variables $z_{(1)}(k)\in R^{12}$, the elements $z_i(k)$ are expressed as
\begin{equation}
    z_i(k) = \psi_i(X(k))=X_i(k),\,i=1,2,\dots,12.
\end{equation}
Utility functions components $z_{(2)}(k)\in R^{18}$ are expressed as  
\begin{equation}
    z_{(2)} = \psi_{(2)} (X(k)) = [\psi_{13} (X(k)),\dots, \psi_{30} (X(k))]^T.
\end{equation}

\bigskip
\textbf{Linear model identification}. To estimated the linear model of the nonlinear 3-robot system (\ref{y(k)_linear_paraed}), as in (\ref{epsilon(k)})-(\ref{Pk1}), we have introduced

\begin{equation}
    \hat{\Theta} (k)= [\hat{A}_{21}(k),\hat{A}_{22}(k),\hat{B}_{2}(k)]^T\in R^{36\times 18},
\end{equation}
\begin{equation}
    \zeta(k) = [z_{(1)}^T(k), z_{(2)}^T(k), U^T(k)]^T \in R^{36}.    
\end{equation}

The iterative algorithm for $\hat{\Theta}(k)$ as the estimate of $\Theta$ in (\ref{y(k)_linear_paraed}) has been given in (\ref{epsilon(k)})-(\ref{Pk1}).

The natural state variables $z_{(1)}(k+1) = [z_1(k+1),\dots,z_{12}(k+1)]^T=X(k+1) \in R ^{12}$ for the iterative algorithm is based on 
\beq
    X(k+1) = A X(k) + B U(k),
\eeq
where $U(k)$ is a chosen input vector, and the utility functions components $z_{(2)}(k+1)\in R^{18}$ are updated from
\beq 
z_{(2)} (k+1) = \psi_{(2)}(X(k+1)).
\label{z2nonlinear}
\eeq

\bigskip
\textbf{Error signals}. This report considers two  error signals mainly. 

\begin{itemize}
    \item To estimated the unknown parameter matrix $\Theta$, we define the estimation error $\epsilon(k)$:
    \begin{equation}
    \begin{split}
        \epsilon(k) = \; & \hat{\Theta}^T(k) \zeta(k)-y(k)\\
    =& \; \hat{\Theta}^T(k)\zeta(k) - z_{(2)}(k+1)\\
    =& \;\hat{A}_{21}(k) z_{(1)}(k) + \hat{A}_{22}(k) z_{(2)}(k) + \hat{B}_2(k)U(k)- z_{(2)}(k+1). 
    \end{split}
    \label{estimationerror}
    \end{equation}

    \item To check the performance of the parameter estimation algorithm, we define the \textit{posteriori} estimation error $\epsilon_a(k)$: 
    \begin{equation}
        \begin{split}
            \epsilon_a(k) = \; & \hat{\Theta}^T(k+1) \zeta(k)-y(k)\\
    =& \; \hat{\Theta}^T(k+1)\zeta(k) - z_{(2)}(k+1)\\
    =& \;\hat{A}_{21}(k+1) z_{(1)}(k) + \hat{A}_{22}(k+1) z_{(2)}(k) + \hat{B}_2(k+1)U(k)- z_{(2)}(k+1). 
        \end{split}
        \label{p_estimationerror}
    \end{equation}
The posteriori estimation error $\epsilon_a(k)$ should converge to $\epsilon(k)$ once the estimated matrix $\hat{\Theta}(k)$ can converge to a constant matrix. In this case, the posteriori estimation error $\epsilon_a(k)$ is more proper to demonstrate the parameter estimation algorithm performance since the $\hat{\Theta}(k)$ may not converge to a constant matrix. 
\end{itemize}

\begin{figure}[H]
    \begin{center}
\includegraphics[width=0.7\textwidth]{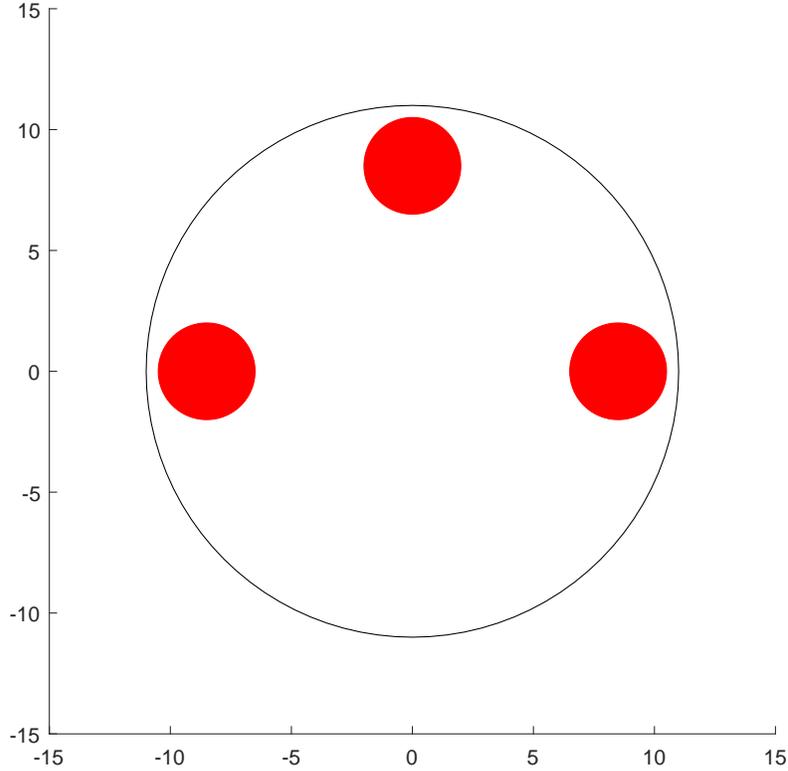}
\caption{3-Robot system work space schematic diagram.}
\label{workspace}
\end{center}
\end{figure}

\bigskip
\textbf{Remark}. The work space of this system is shown in Figure \ref{workspace}, where the largest black border circle is a circular wall ($radius = 11 m$), and the three red circles are the three movable disk-like robots ($radius = 2 m$). 
 In the processes of linear system identifications and model validations, the collisions between robots with robots and robots with the circular wall should be avoided. 

\subsubsection{Simulation Results}

In the simulation, we set the sampling time $\Delta t = 0.05 $ s, $\rho=0.000 1$ as the iterative algorithm design parameter, $P_0 = P_0^T = 100I_{36} >0$\footnote{The matrix $I_{36}$ denotes the identity matrix of size $36\times36$. } and $\Theta_0=0_{36\times 18}$ as the initial values of the iterative algorithm. The initial value of the robot state vector $X(0)=z_{(1)}(0) = [x_1(0),y_1(0),\dots ,x_3(0),y_3(0),v^x_1(0),v^y_1(0), \\ \dots ,v^x_3(0),v^y_3(0)]^T = [-7,3,0,7,7,-4,0,0,0,0,0,0]^T $, which means Robot 1, Robot 2, and, Robot 3 are located at $(-7,3)$, $(0,7)$, $(7,-4)$ with zero initial speeds in X-direction and Y-direction at $k=0$. Correspondingly, the target positions of the three robots are $(-4.5,0)$, $(3, 4)$, and $(3,-4)$. In $k_s=300$ iterations, apply a selected input vector $U(k) = [a_1^x,a_1^y,a_2^x,a_2^y,a_3^x,a_3^y]^T$ (as shown in Table \ref{S1_MI_in}) to do the linear model approximation according to \cite{t21},\cite{zt22}. 
\bigskip
\par \textbf{Matrix $P(k)$ resetting.} To improve the estimation results, we reset the matrix $P(k)=P_0$ every 130 iterations.

\begin{figure}[H]
\begin{center}
\includegraphics[width=1\textwidth]{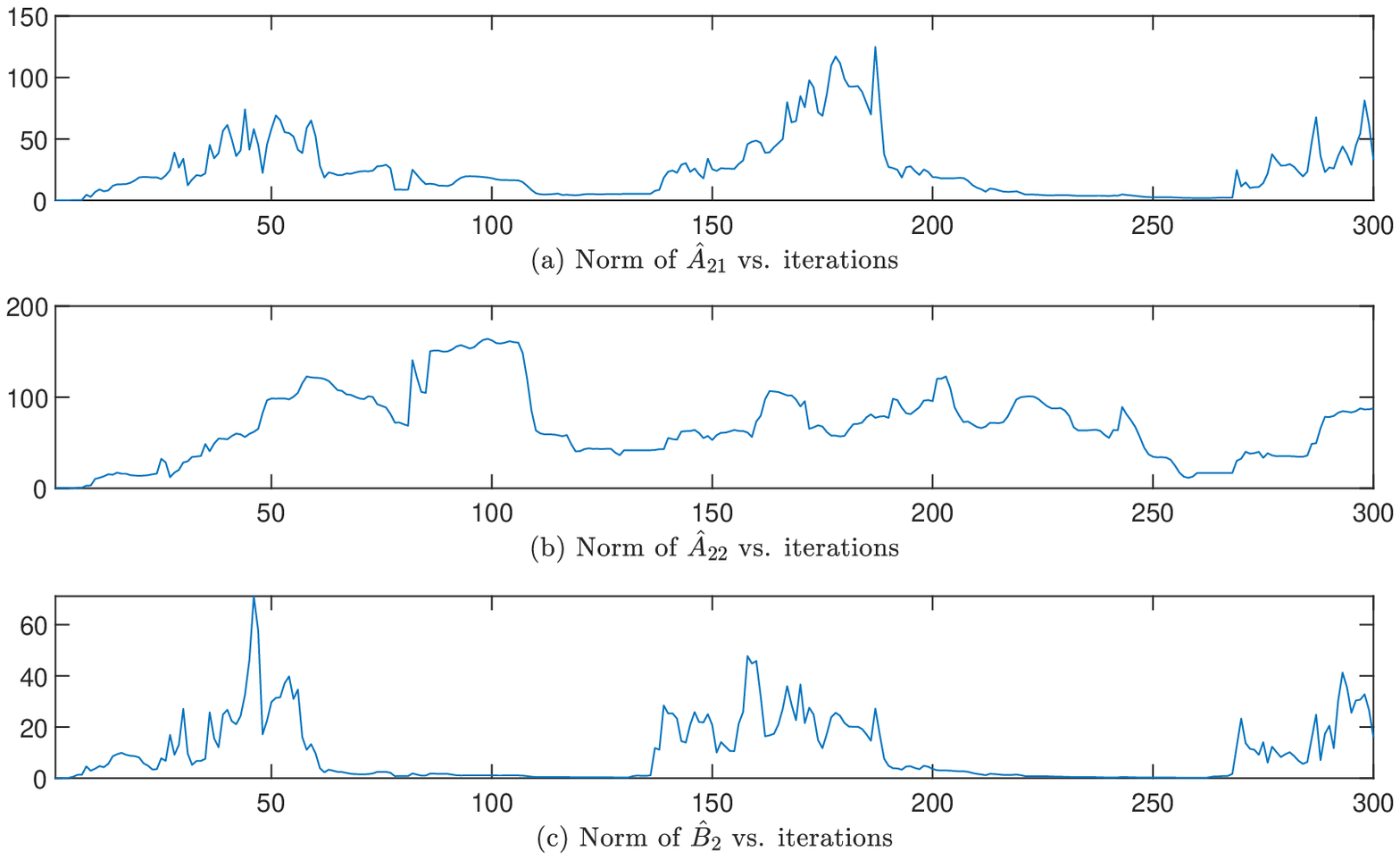}
\caption{Norms of $\hat{\Theta}(k) $ (linear model identification for 3-robot system).}
\label{3r_linear_theta_norm}
\end{center}
\end{figure}

\begin{figure}[H]
\begin{center}
\includegraphics[width=0.78\textwidth]{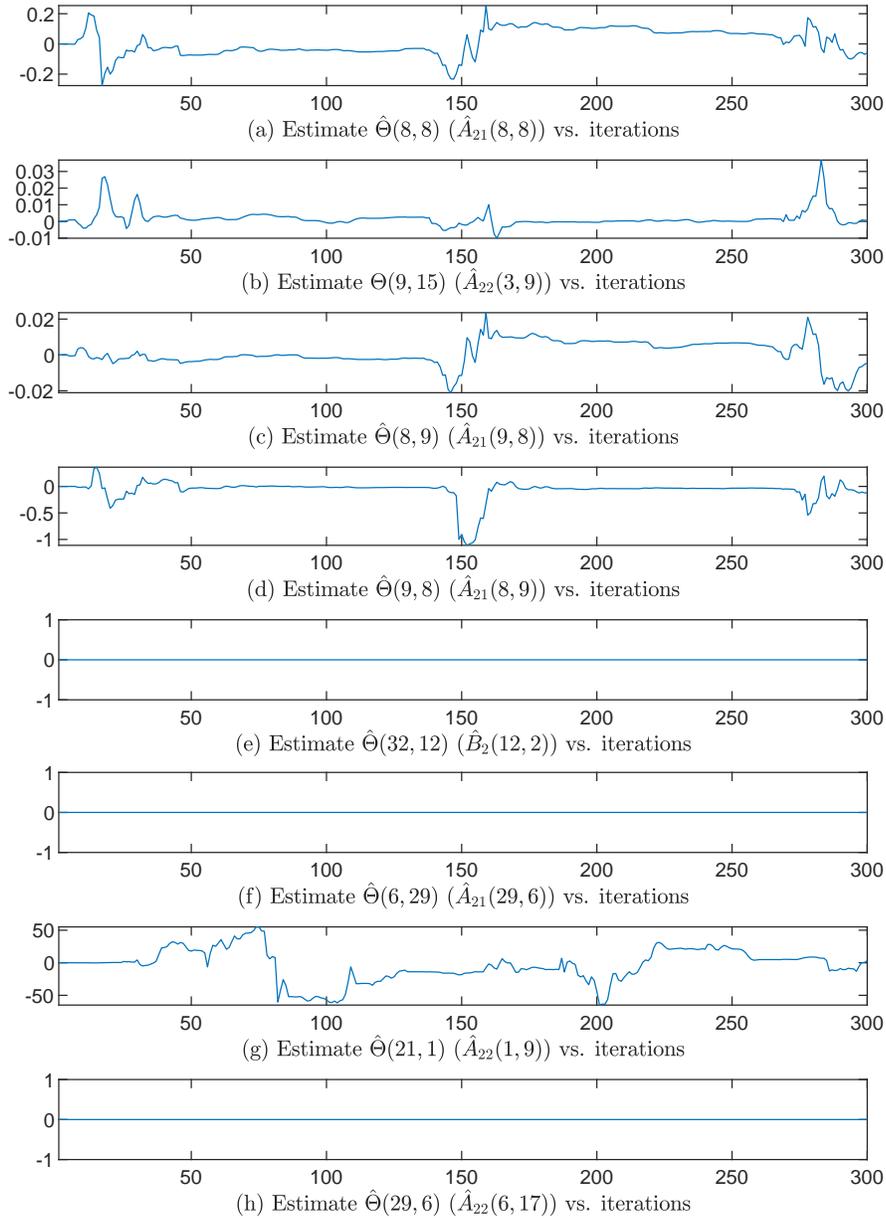}
\caption{Components of $\hat{\Theta}(k) $ (linear model identification for 3-robot system)\protect\footnotemark.}
\label{3r_linear_theta_com}
\end{center}
\end{figure}

\begin{table}[H]
\renewcommand\arraystretch{1.5}

\centering
\begin{tabular}{cc}

\hline
\hline
\multicolumn{2}{c}{System Inputs} \\
\hline
Input & Signal\\
\hline
$a_1^x$ & $2\sin(0.075{k\pi })$\\
$a_1^y$ & $3\cos(0.1833{k\pi  })$\\
$a_2^x$ & $3\sin(0.14{k\pi  })$\\ 
$a_2^y$ & $3\cos(0.095{k\pi  })$\\
$a_3^x$ & $-2\sin(0.06{k\pi })$\\ 
$a_3^y$ & $-3\cos(0.092{k\pi })$\\[1ex]
\hline
\hline

\end{tabular}
\caption{Model identification system input components.}

\label{S1_MI_in}
\end{table}
{\bf Parameter estimates}. Figures \ref{3r_linear_theta_norm}-\ref{3r_linear_theta_com} shows the results of parameter estimates. 

\newpage

\footnotetext{In
this figure, the notation $\hat{\Theta}(8,8)$ denotes the $(8,8)$th
component of the matrix $\hat{\Theta}(k) = [\hat{\Theta}_{ij}(k)]$
whose $(i,j)$th components are $\hat{\Theta}_{ij}(k)$. The
notation $\hat{A}_{21}(8,8)$ denotes the $(8,8)$th component of the
matrix $\hat{A}_{21}(k)$. Other notations have similar meanings.}

\begin{figure}[H]
\begin{center}

\includegraphics[width=0.85\textwidth]{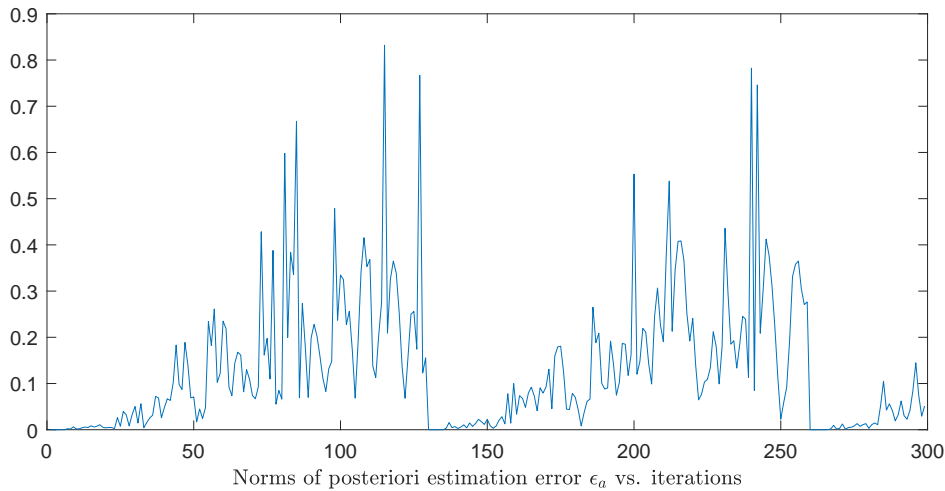}
\caption{Norm of the \textit{posteriori} estimation error $\epsilon_a(k) $ (linear model based).}
\label{Control_eps_norm_linear}

\end{center}
\end{figure}

\bigskip
{\bf Estimation errors}. The results of the \textit{posteriori} estimation error introduced in (\ref{p_estimationerror}), $
\epsilon_a(k) =  \hat{\Theta}(k+1)\zeta(k)  - z_{(2)}(k+1)$
are presented in Figure \ref{Control_eps_norm_linear}. 

\begin{figure}[H]
\begin{center}

\includegraphics[width=0.85\textwidth]{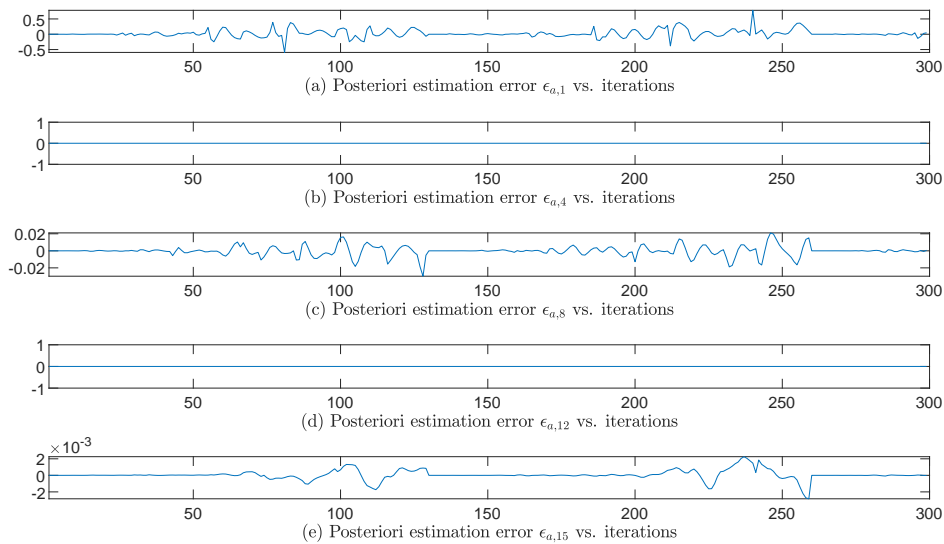}
\caption{Components of the \textit{posteriori} estimation error $\epsilon_a(k) $ (linear model based)\protect\footnotemark.}
\label{Control_eps_com_linear}
\end{center}
\end{figure}
\footnotetext{In
this figure, the notation ${\epsilon_{a,1}}$ denotes the first
component of the vector ${\epsilon_a}(k) $. Other notations have similar meanings.}

{\bf Discussion}. The parameter matrix norms and some selected elements of the parameter matrix are demonstrated in Figures \ref{3r_linear_theta_norm}-\ref{3r_linear_theta_com}. The results of the \textit{posteriori} estimation error introduced in (\ref{p_estimationerror}), $
\epsilon_a(k) =  \hat{\Theta}(k+1)\zeta(k)-z_{(2)}(k+1)$
are presented in Figure \ref{Control_eps_norm_linear}. 
 From Figures \ref{Control_eps_norm_linear}-\ref{Control_eps_com_linear}, we find that our iterative algorithm cannot estimate the parameters of the linear model well although both the \textit{posteriori} estimation error norm and components are quite low. However, there is a non-ideal phenomenon in the simulation, the \textit{posteriori} estimation error and the parameter estimates always have an obvious periodic increment tendency (from 0 to almost 0.9) during the iterations between two adjacent times we reset the parameter matrix $P$. This indicates that the estimated model cannot always ensure the estimation error within a small range, which may lead to undesired results while using the linear model based control signals. 

\bigskip
\textbf{Remark}. In the above simulation, we also consider the distances between robots and the circular wall to check whether there are collisions happening. Since all the distances are larger than zero, no collisions happen during the identification process. 
\par In \cite{zt22}, the linear model approximation simulation is implemented in a 5-robot system. According to the results, both the model validation error (the difference between the nonlinear utility component vector and the estimated output from the linear model with the fixed parameter matrix $\hat{\Theta}$ generated by several iterations of the adaptive parameter estimation algorithm) and the estimation error (the difference between the nonlinear utility component vector and the estimated output from the linear model with the online updating matrix $\hat{\Theta}(k)$ generated by minimizing the cost function $J_3$, defined in (\ref{J_3}), of the adaptive parameter estimation algorithm) are quite low. However, tendencies of periodic increases in the estimation errors also appear in the 5-robot system simulation cases. This further proves the linear model may not be powerful to approximate the nonlinear utility functions and cannot be suitable to solve the optimal control problem.

\subsection{Bilinear Model Identification of 3-robot System}
\label{Sec_simuBilinID}
In this subsection, the simulation system and the simulation results for the bilinear model identification of the 3-robot system are demonstrated sequentially. The estimation errors are examined to check the performance of the model (system) identification. The selected parameter components and the norms of the parameter matrices are analyzed for the parameter results.

\subsubsection{Simulation System}
The basic simulation construction and data collection logic are introduced in the following. 

\bigskip
\textbf{System state variables}. System state variables $z(k)$ are constructed with natural state variables and utility function components. For the natural state variables $z_{(1)}(k)\in R^{12}$, the elements $z_i(k)$ are expressed as
\begin{equation}
    z_i(k) = \psi_i(X(k))=X_i(k),\,i=1,2,\dots,12.
\end{equation}
Utility functions components $z_{(2)}(k)\in R^{18}$ are expressed as  
\begin{equation}
    z_{(2)} = \psi_{(2)} (X(k)) = [\psi_{13} (X(k)),\dots, \psi_{30} (X(k))]^T.
\end{equation}

\bigskip
\textbf{Bilinear model identification}. To estimated the linear model of the nonlinear 3-robot system (\ref{bilinear_model}), as in (\ref{y(k)(z_2)_bilinear_raw})-(\ref{z(2)(k+1)appr_bilinear}), we have introduced

\begin{equation}
    \hat{\Theta} (k)= [\hat{A}_{21}(k),\hat{A}_{22}(k),\hat{B}_{2}(k), \hat{f}(z_{(1)0},\,z_{(2)0}),\hat{\Phi}_1^T(k),\hat{\Phi}_2^T(k),\hat{\Phi}_3^T(k),\hat{\Phi}_4^T(k),\hat{\Phi}_5^T(k)]^T\in R^{901\times18},
\end{equation}
\begin{equation}
    \zeta(k) = [(z_{(1)}(k)-z_{(1)0})^T,(z_{(2)}(k)-z_{(2)0})^T,U^T(k), 1, g_1^T(\cdot),g_2^T(\cdot,\cdot),g_3^T(\cdot),g_4^T(\cdot,\cdot),g_5^T(\cdot,\cdot)]^T \in R^{901} .
    \label{Thetaaa}
\end{equation}

The iterative algorithm for $\hat{\Theta}(k) = [\hat{\theta}_1(k),\hat{\theta}_2(k),\dots,\hat{\theta}_{18}(k)] \in R^{901\times18}$ as the estimate of $\Theta$ in (\ref{y(k)_bilinear_paraed}) has been given in (\ref{Thetak+1})-(\ref{Pk1}).

\bigskip
\textbf{Error signals}. Similar to the linear model identification simulation, we also have two kinds of different error signals here:

\begin{itemize}
    \item The estimation error $\epsilon(k) = \hat{\Theta}^T(k)\zeta(k) - z_{(2)}(k+1)$
    is defined to estimate the unknown parameter matrix $\Theta$. 
    
    \item The \textit{posteriori} estimation error $\epsilon_a(k) =  \hat{\Theta}^T(k+1)\zeta(k) - z_{(2)}(k+1)$
    is defined to check the parameter estimation algorithm performance. The \textit{posteriori} estimation error $\epsilon_a(k)$ is more meaningful for the algorithm performance because the parameter estimate $\hat{\Theta}(k)$ may not converge. 
    
\end{itemize}

\subsubsection{Simulation Results}
In the simulation, we set the sampling time $\Delta t = 0.05 $ s, $\rho=0.000 1$ as the iterative algorithm design parameter, $P_0 = P_0^T = 100I_{901} >0$\footnote{The matrix $I_{901}$ denotes the identity matrix of size $901$. } and $\Theta_0=0_{901\times 18}$ as the initial values of the iterative algorithm. The initial value of the robot state vector $X(0)=z_{(1)}(0) = [x_1(0),y_1(0),\dots ,x_3(0),y_3(0),v^x_1(0),v^y_1(0), \\ \dots ,v^x_3(0),v^y_3(0)]^T = [-7,3,0,7,7,-4,0,0,0,0,0,0]^T $, which means Robot 1, Robot 2, and, Robot 3 are located at $(-7,3)$, $(0,7)$, $(7,-4)$ with zero initial speeds in X-direction and Y-direction at $k=0$. Correspondingly, the target positions of the three robots are $(-4.5,0)$, $(3, 4)$, and $(3,-4)$. In $k_s=300$ iterations, apply a selected input vector (as shown in Table \ref{S1_MI_in}) to do the linear model approximation with the simulation system introduced above. 

\bigskip
\par \textbf{Matrix $P(k)$ resetting.} To improve the estimation results, we reset the matrix $P(k)=P_0$ every 75 iterations.

\begin{figure}[H]
\begin{center}
\includegraphics[width=.98\textwidth]{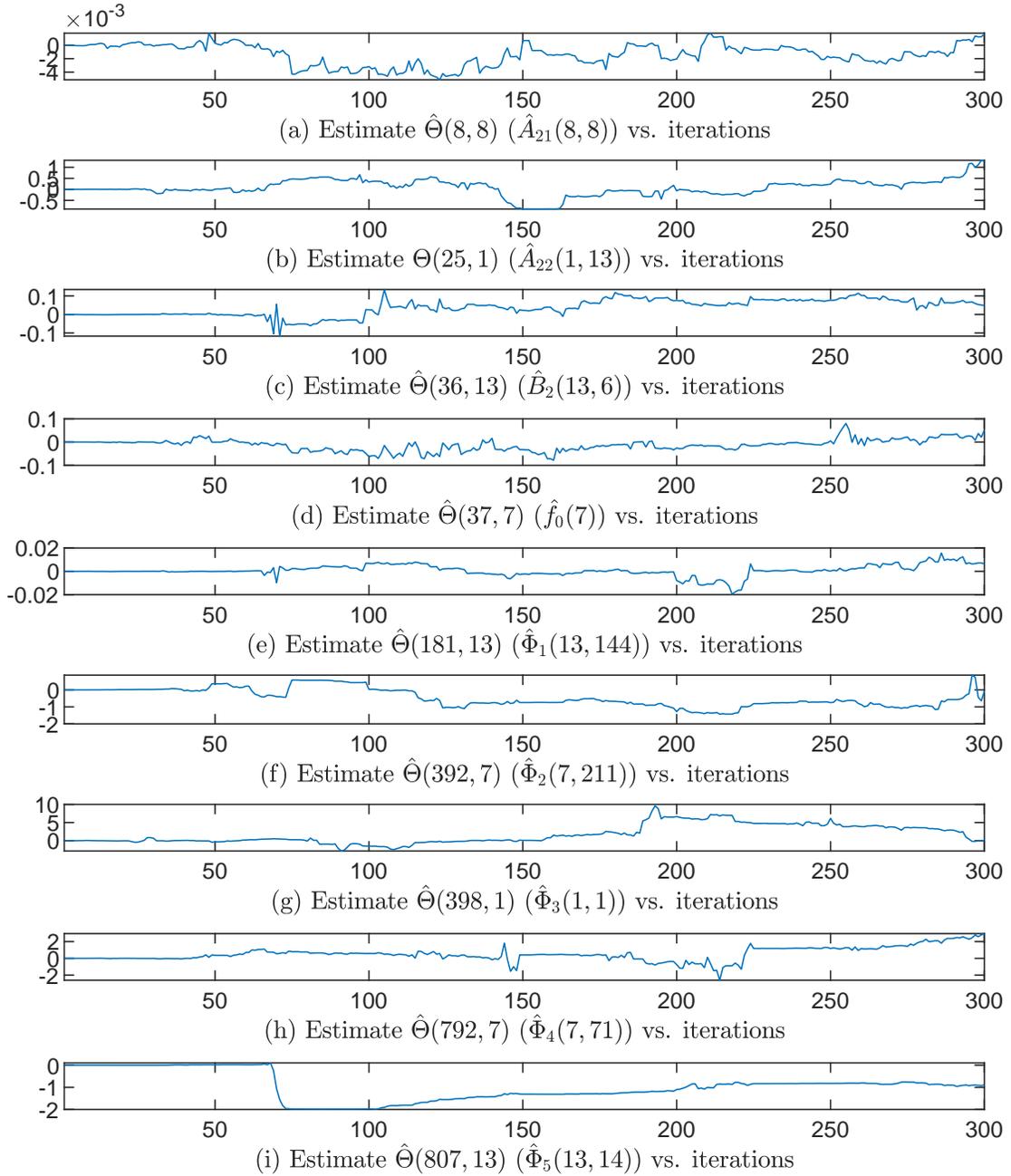}
\caption{Components of $\hat{\Theta}(k) $ (bilinear model identification for 3-robot system).}
\label{S3_Theta_re2.ele}
\end{center}
\end{figure}

\begin{figure}[H]
\begin{center}
\includegraphics[width=1\textwidth]{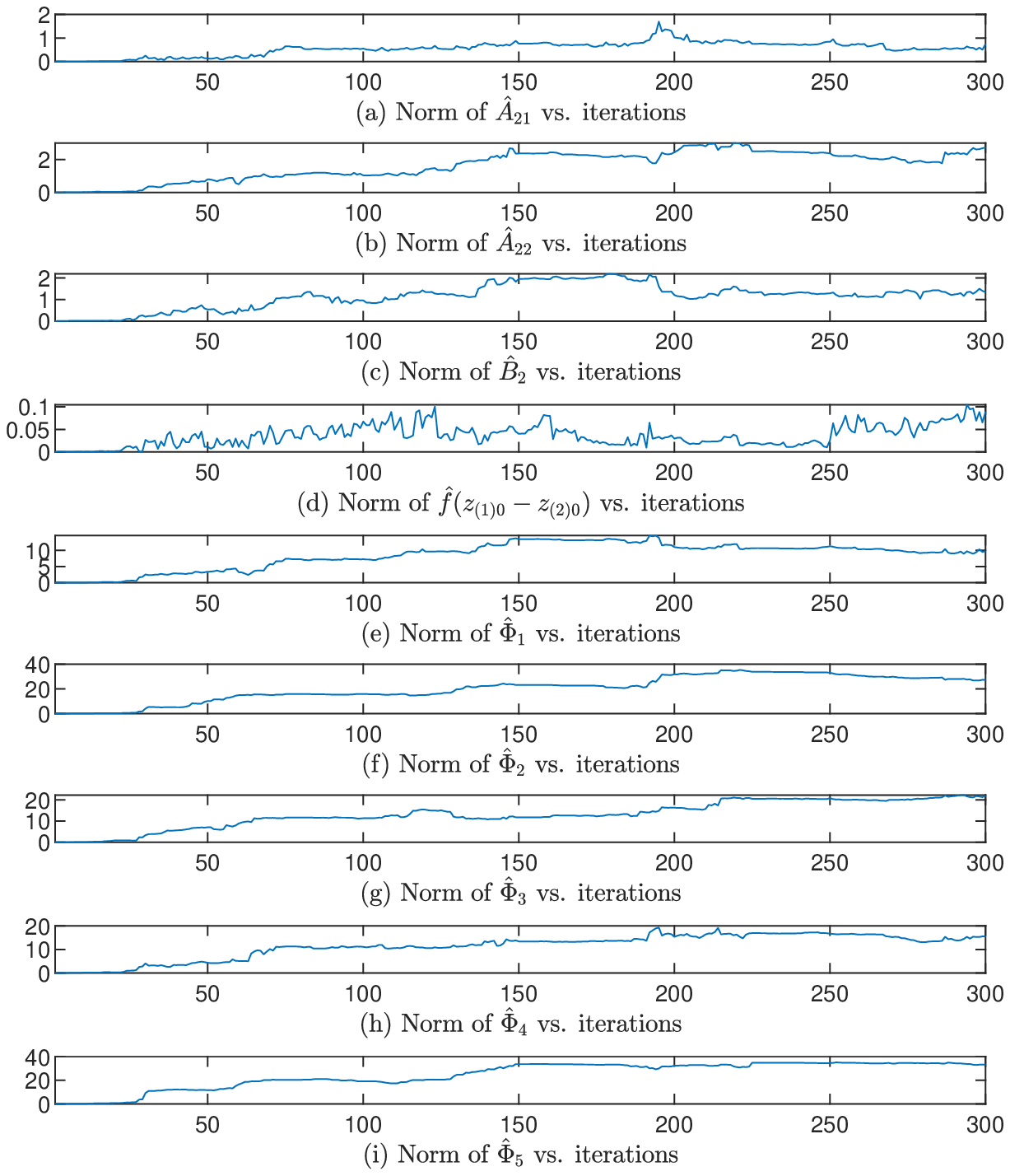}
\caption{Norms of $\hat{\Theta}(k) $ (bilinear model identification for 3-robot system).}
\label{S3_Theta_re2.norm}
\end{center}
\end{figure}

\newpage
{\bf Parameter estimates}. Figures \ref{S3_Theta_re2.ele}-\ref{S3_Theta_re2.norm} show the results of parameter estimates. 

\begin{figure}[H]
\begin{center}

\includegraphics[width = 0.95 \textwidth]{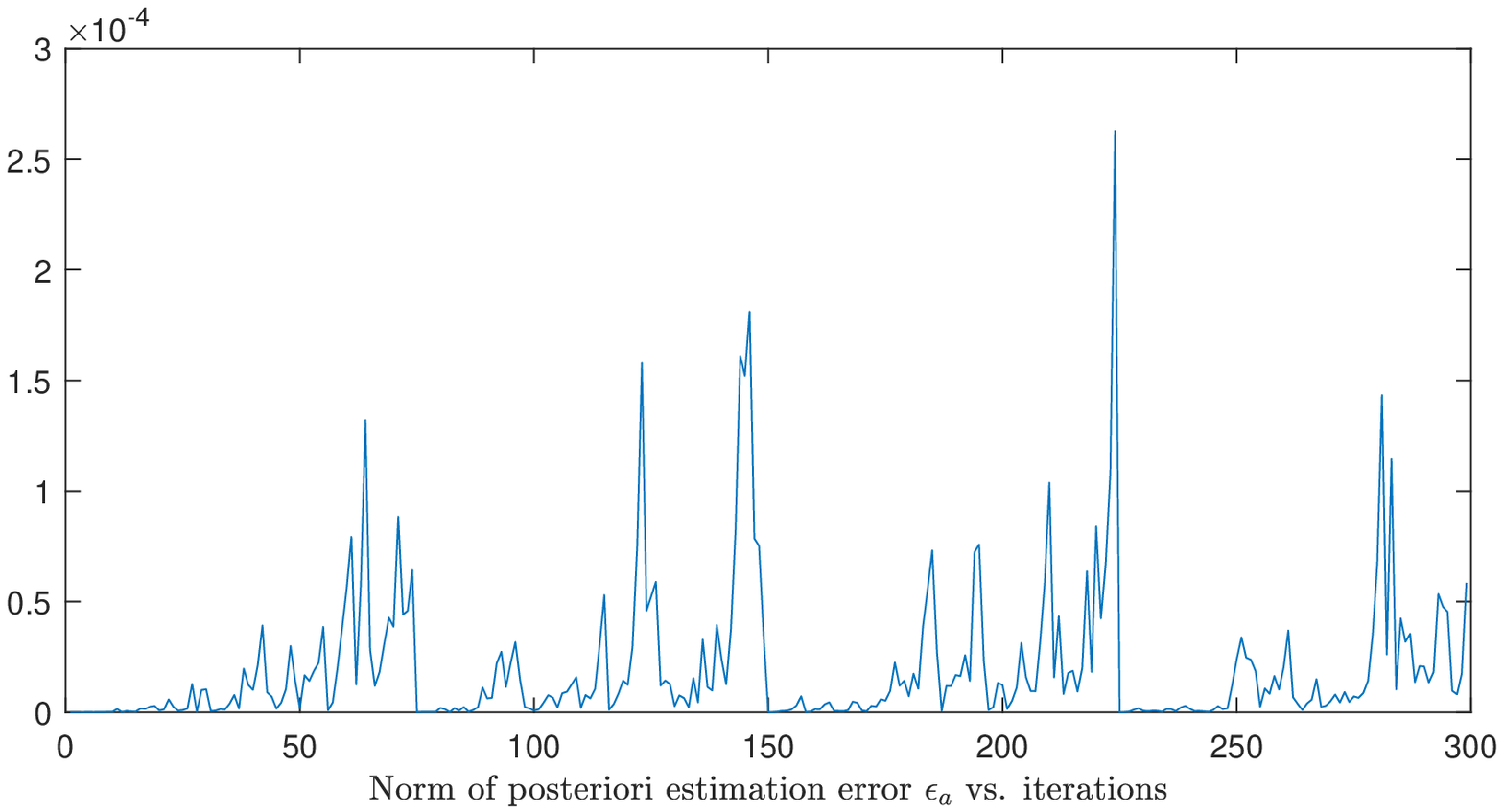}
\caption{Norm of the \textit{posteriori} estimation error $\epsilon_a(k) $ (bilinear model based).}
\label{S1_eps.norm}

\end{center}
\end{figure}

\begin{figure}[H]
\begin{center}

\includegraphics[width=0.95\textwidth]{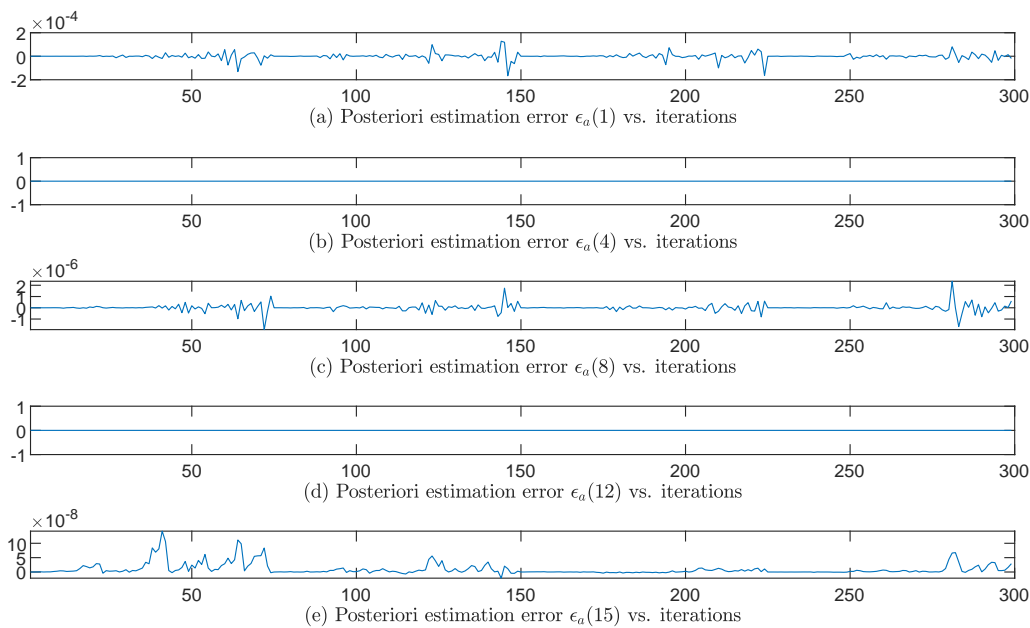}
\caption{Components of the \textit{posteriori} estimation error $\epsilon_a(k) $ (bilinear model based).}
\label{S1_eps.com}
\end{center}
\end{figure}

\newpage
{\bf Estimation errors}. The results of the \textit{posteriori} estimation error introduced in (\ref{estimationerror}), $\epsilon_a(k) = \hat{\Theta}(k+1)\zeta(k)-z_{(2)}(k+1)$
are presented in Figures \ref{S1_eps.norm}-\ref{S1_eps.com}.

\bigskip

\textbf{Remark}. In the simulation, all the distances between robots and the circular wall are larger than zero, meaning no collisions happen between the robot and others or between the robot and the circular wall during this process.

\bigskip
{\bf Discussion}. From Figures \ref{S1_eps.norm}-\ref{S1_eps.com}, we can find that our iterative algorithm can estimate the parameters of the bilinear model well since the norm of the \textit{posteriori} estimation error $\epsilon_a(k)$ is always less than $3\times 10 ^{-4}$ and the components of the \textit{posteriori} estimation error $\epsilon_a(k)$ are also quite small. Figures \ref{S3_Theta_re2.ele}-\ref{S3_Theta_re2.norm} indicate that all the parameter estimate norms and selected components change within a quite small range, although they do not converge. Thus, the bilinear model can better approximate the 3-robot system with the nonlinear utility functions since the \textit{posteriori} estimation error of the bilinear model is much smaller than the \textit{posteriori} estimation error of the linear model. Therefore, this bilinear model has a better performance to approximate the nonlinear utility functions although the tendency for periodic changes also appears in the corresponding results of the estimation error from the bilinear approximation model.

\subsection{Linear and Bilinear Model Based Optimal Control for the 3-Robot System}
\label{Sec_simuCE}
In this subsection, the simulation system and the simulation results for evaluating the linear model based control design and the bilinear model based control design of the 3-robot system are demonstrated sequentially. The robot trajectories are presented to check whether the desired tracking objectives are achieved. During the feedback control phase, the distances between robots with others and robots with the wall are analyzed to check whether collisions happen.

\bigskip
\subsubsection{Simulation Conditions}
Before testing the performance of the proposed baseline control input, we need to use some iterations to estimate the linear approximation for the nonlinear model. Thus, there are two phases in the simulations for the baseline control input:
\begin{itemize}
    \item \textbf{Phase I (System Identification Phase)}: Initially, we set the sampling time $\Delta t = 0.05 $ s, $\rho=0.000 1$ as the iterative algorithm design parameter, $P_0 = P_0^T = 100I_{a} >0$ and $\Theta_0=0_{a\times 18}$ as the initial values of the iterative algorithm\footnote{The values of $a$ are vary for different models. For the linear model, the value of $a$ is $36$. For the bilinear model, the value of $a$ is $901$.  }\footnote{The matrix $I_{a}$ denotes the identity matrix of size $a$. }. The initial value of the robot state vector $X(0)=z_{(1)}(0) = [x_1(0),y_1(0),\dots ,x_3(0),y_3(0),v^x_1(0),v^y_1(0),\dots ,v^x_3(0),v^y_3(0)]^T = [-7,3,0,7,7,-4,0,0,0,0,0,0]^T $, which means Robot 1, Robot 2, and, Robot 3 are located at $(-7,3)$, $(0,7)$, $(7,-4)$ with zero initial speeds in X-direction and Y-direction at $k=0$. Correspondingly, the target positions of the three robots are $(-4.5,0)$, $(3, 4)$, and $(3,-4)$. In the first $k_s=300$ iterations, apply a selected proper input vector to do the linear or bilinear model approximation according to \cite{t21},\cite{zt22}. The detailed steps for each system identification iteration are listed in the following:
    \begin{enumerate}
        \item [\textbf{Step 1}: ] Compute the input $U(k)$, whose components are listed in Table \ref{S1_MI_in}. 
        \item [\textbf{Step 2}: ] For the linear model identification, update the robots state $z_{(1)}(k)$ based on the robots dynamic equation (\ref{3r_z1_update}). For the bilinear model identification, update the vector $z_{(1)}(k)-z_{(1)0}$ according to (\ref{3r_z1_update}) and (\ref{x0}). 
\item [\textbf{Step 3}: ] For the linear model identification, calculate the utility components $z_{(2)}(k)$ based on equations (\ref{phi_c1})-(\ref{phi_c6}) and (\ref{zig2}). For the bilinear model identification, calculate the vector $z_{(2)}(k)-z_{(2)0}$ based on equations (\ref{phi_c1})-(\ref{phi_c6}), (\ref{zig2}) and (\ref{x0}).

\item [\textbf{Step 4}: ] Update the approximations $\hat{z}_{(2)}(k+1) = \hat{\Theta}^T(k)\zeta(k) $, with $\hat{\Theta}^T(k) $, $\zeta(k) $ for the nonlinear utility functions based on equations (\ref{epsilon(k)})-(\ref{Pk1}). For the linear model identification, the signal vector $\zeta(k) $ and the parameter matrix $\hat{\Theta}(k) $ are defined in equations (\ref{zeta}) and (\ref{theta_hat_linear}). Correspondingly, the the signal vector $\zeta(k) $ and the parameter matrix $\hat{\Theta}(k) $ of the bilinear model identification are defined in equations (\ref{zeta_bilinear}) and (\ref{theta_hat_bilinear}). 
    \end{enumerate}
    \par \textbf{Matrix $P(k)$ resetting.}  To improve the estimation results, we reset the matrix $P(k)=P_0$ every 130 iterations for linear model identification cases. Similarly, the matrix $P(k)$ is reset to be $P_0$ every 75 iterations in the bilinear model identification cases. 

    \item \textbf{Phase II (Feedback Control Phase)}: From the $(k_s+1) = 301 $th iteration, reset the initial and target positions of the robots and replace the input vector with the results of the linear programming problem to test the baseline control input performance and keep the updating of the model approximation results. Table \ref{robot_setting_bi} demonstrates the detailed robot initial and target position settings of the selected three cases\footnote{Since the linear approximation model does not contain enough information to estimate the nonlinear utility functions of the 3-robot system, the linear model based control design cannot achieve the objectives. Thus, we only present the result of the linear model based control designs in Case I here to save space.}. The details are listed in the following:

 \begin{enumerate}
        \item [\textbf{Step 1}: ]Generate the input $U(k)$, which is the solution for the linear programming problem shown in equation (\ref{sol_inputfromlinear}) or (\ref{sol_inputfrombilinear}). 
        
       \item [\textbf{Step 2}: ] For the linear model identification, update the robots state $z_{(1)}(k)$ based on the robots dynamic equation (\ref{3r_z1_update}). For the bilinear model identification, update the vector $z_{(1)}(k)-z_{(1)0}$ according to (\ref{3r_z1_update}) and (\ref{x0}). 
\item [\textbf{Step 3}: ] For the linear model identification, calculate the utility components $z_{(2)}(k)$ based on equations (\ref{phi_c1})-(\ref{phi_c6}) and (\ref{zig2}). For the bilinear model identification, calculate the vector $z_{(2)}(k)-z_{(2)0}$ based on equations (\ref{phi_c1})-(\ref{phi_c6}), (\ref{zig2}) and (\ref{x0}).

\item [\textbf{Step 4}: ] Update the approximations $\hat{z}_{(2)}(k+1) = \hat{\Theta}^T(k)\zeta(k) $, with $\hat{\Theta}^T(k) $, $\zeta(k) $ for the nonlinear utility functions based on equations (\ref{epsilon(k)})-(\ref{Pk1}). For the linear model identification, the signal vector $\zeta(k) $ and the parameter matrix $\hat{\Theta}(k) $ are defined in equations (\ref{zeta}) and (\ref{theta_hat_linear}). Correspondingly, the the signal vector $\zeta(k) $ and the parameter matrix $\hat{\Theta}(k) $ of the bilinear model identification are defined in equations (\ref{zeta_bilinear}) and (\ref{theta_hat_bilinear}).

    \end{enumerate}
\begin{table}[H]
    \renewcommand\arraystretch{1.5}

\begin{center}
\begin{tabular}{cccc} 
\hline
\hline
\multicolumn{4}{c}{Case Settings}\\
\hline
Case number             &  Robot number             & Initial position & Target position\\
\hline
\multirow{3}{*}{Case I}              
                        &\multirow{1}{*}{Robot 1}   &      $(-4.48,3.3)$  & $(-4.5,0)$     \\ 
                        &\multirow{1}{*}{Robot 2}   &     $(2.98,7.11)$ & $(3, 4)$     \\ 
                        &\multirow{1}{*}{Robot 3}   &     $(3.65,-5.70)$  & $(3,-4)$   \\

\hline

\multirow{3}{*}{Case II}              
                        &\multirow{1}{*}{Robot 1}   &      $(-3.1,3.3)$&$(-3,0.5)$    \\ 
                        &\multirow{1}{*}{Robot 2}   &     $(4.9,4.6)$&$(5,2)$     \\ 
                        &\multirow{1}{*}{Robot 3}   &     $(0.9,-5)$&$(0.5,-3.5)$   \\ 
\hline   
\multirow{3}{*}{Case III}              
                        &\multirow{1}{*}{Robot 1}   &      $(7.5,2.5)$&$(7.5,0)$    \\ 
                        &\multirow{1}{*}{Robot 2}   &     $(-5.1,-2.5)$&$(-5, -5)$     \\ 
                        &\multirow{1}{*}{Robot 3}   &     $(-5.2,3.5)$&$(-5, 5)$   \\ 

                        \hline   
\multirow{3}{*}{Case IV}              
                        &\multirow{1}{*}{Robot 1}   &      $(-3.1,-0.6)$&$(-3.5,-3.5)$    \\ 
                        &\multirow{1}{*}{Robot 2}   &     $(0, 2.5)$&$(0.2, 0)$     \\ 
                        &\multirow{1}{*}{Robot 3}   &     $(4.8,3)$&$(5.3, 5)$   \\ 
                        \hline
\multirow{3}{*}{Case V}              
                        &\multirow{1}{*}{Robot 1}   &      $(-6.4,5.3)$&$(-6,2.4)$    \\ 
                        &\multirow{1}{*}{Robot 2}   &     $(-2.5,1.8)$&$(-2.5, -1.4)$     \\ 
                        &\multirow{1}{*}{Robot 3}   &     $(3,-2.7)$&$(2.5, -1.4)$   \\ 
                        \hline\hline
\end{tabular}
\end{center}
\caption{Settings of the feedback control phase.}

\label{robot_setting_bi}
\end{table}    
    
\end{itemize}

\subsubsection{Simulation Results}
\begin{figure}[H]
     \centering
     \begin{subfigure}[b]{0.48\textwidth}
         \centering
         \includegraphics[width=0.75\textwidth]{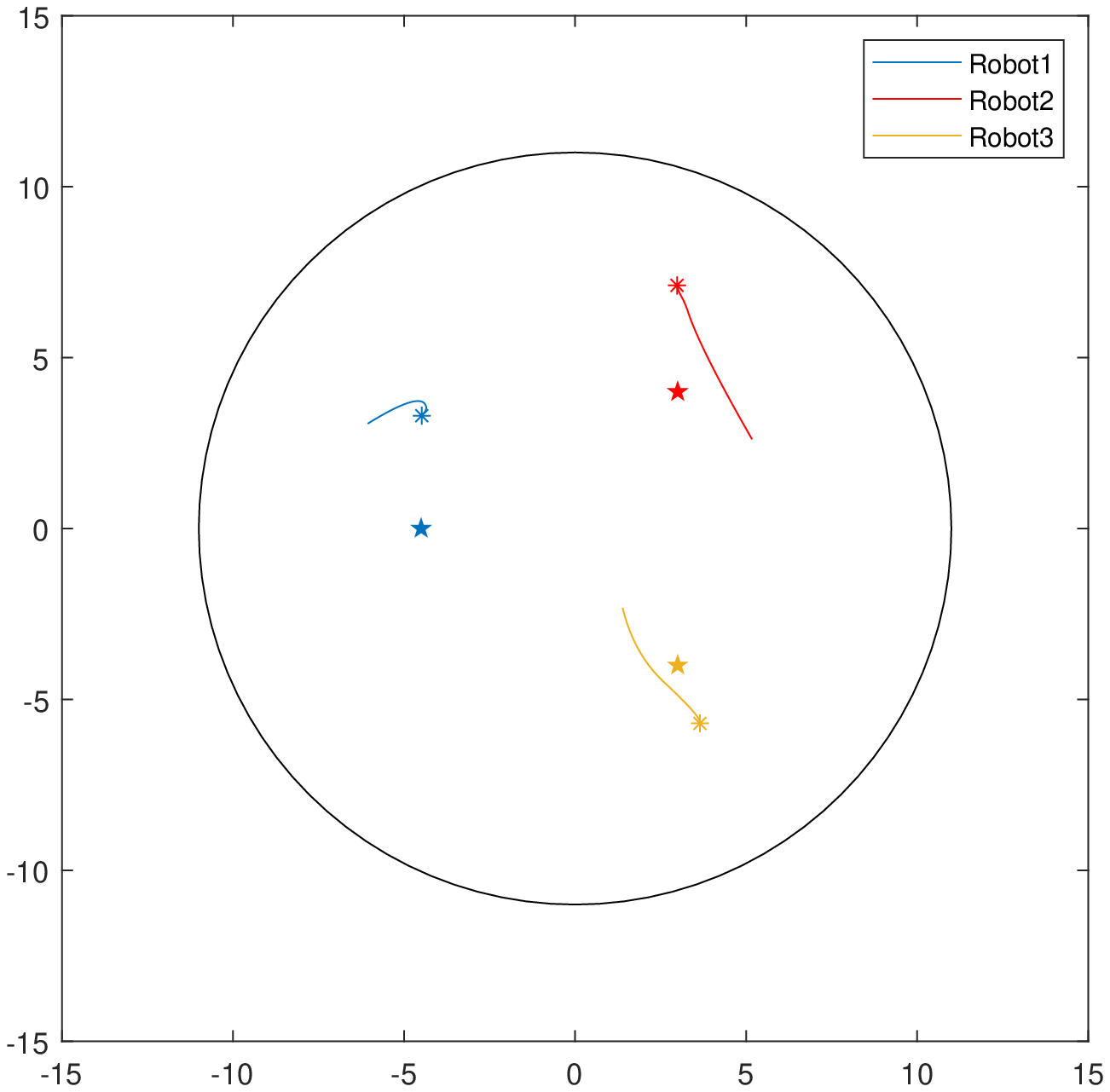}
         \caption{Case I (linear model based control design)}
         \label{Ctrl_eva_1_Linear}
     \end{subfigure}
     \begin{subfigure}[b]{0.48\textwidth}
         \centering
         \includegraphics[width=0.75\textwidth]{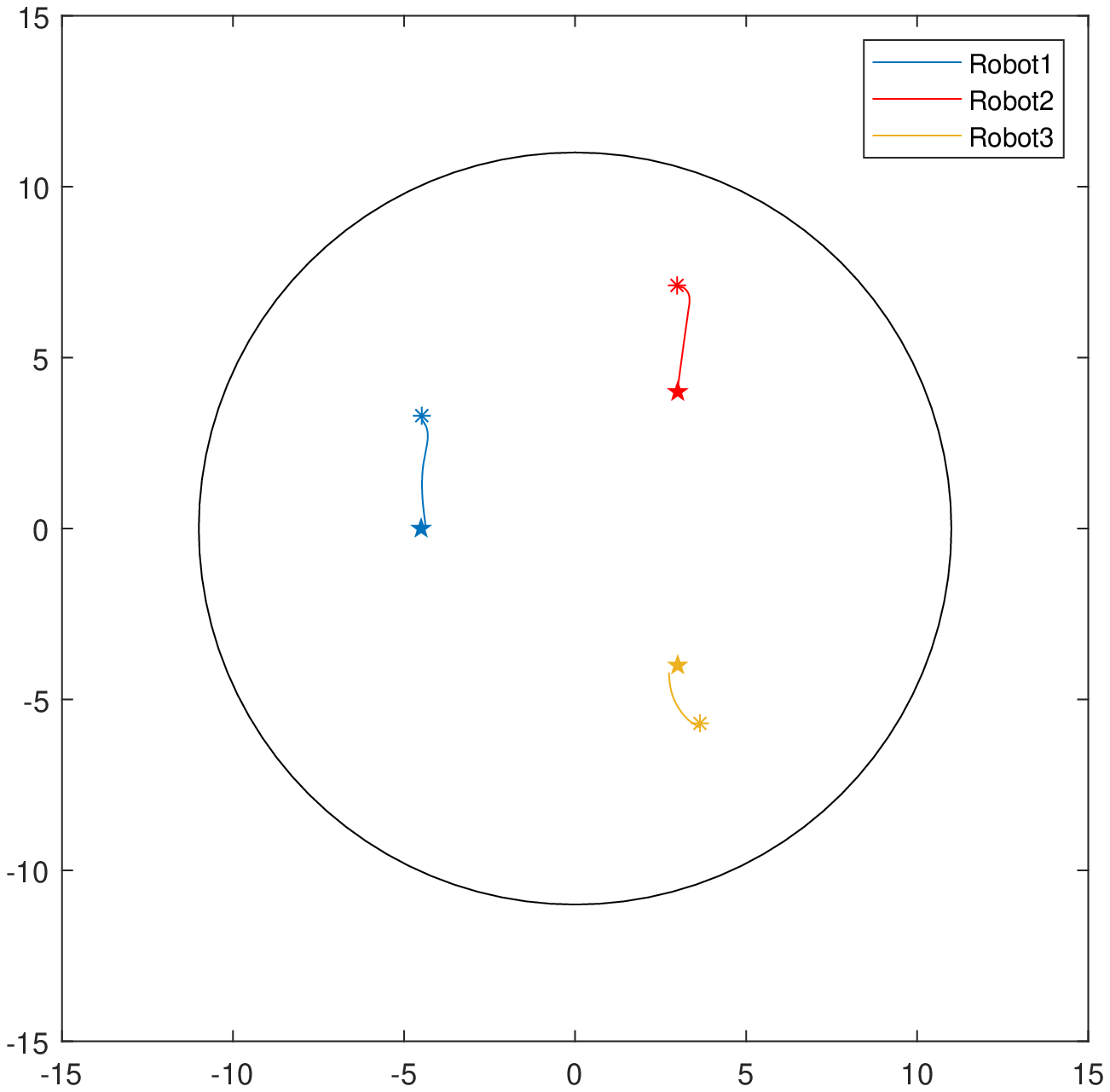}
         \caption{Case I (bilinear model based control design)}
         \label{Ctrl_eva_1_CaseI}
     \end{subfigure}
     \hfill
     \begin{subfigure}[b]{0.48\textwidth}
         \centering
         \includegraphics[width=0.75\textwidth]{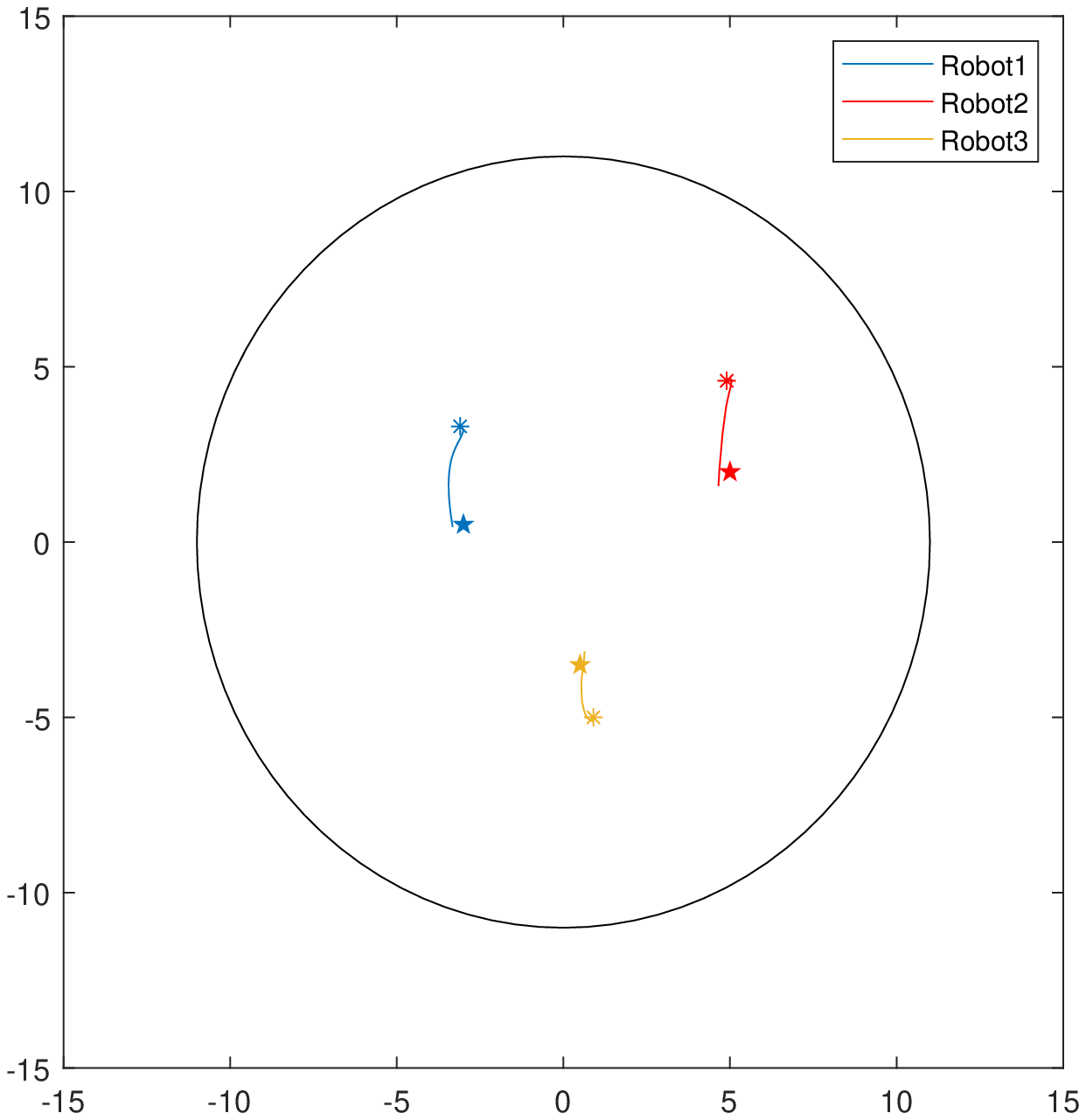}
         \caption{Case II (bilinear model based control design)}
         \label{Ctrl_eva_1_CaseII}
     \end{subfigure}
     \begin{subfigure}[b]{0.48\textwidth}
         \centering
         \includegraphics[width=0.75\textwidth]{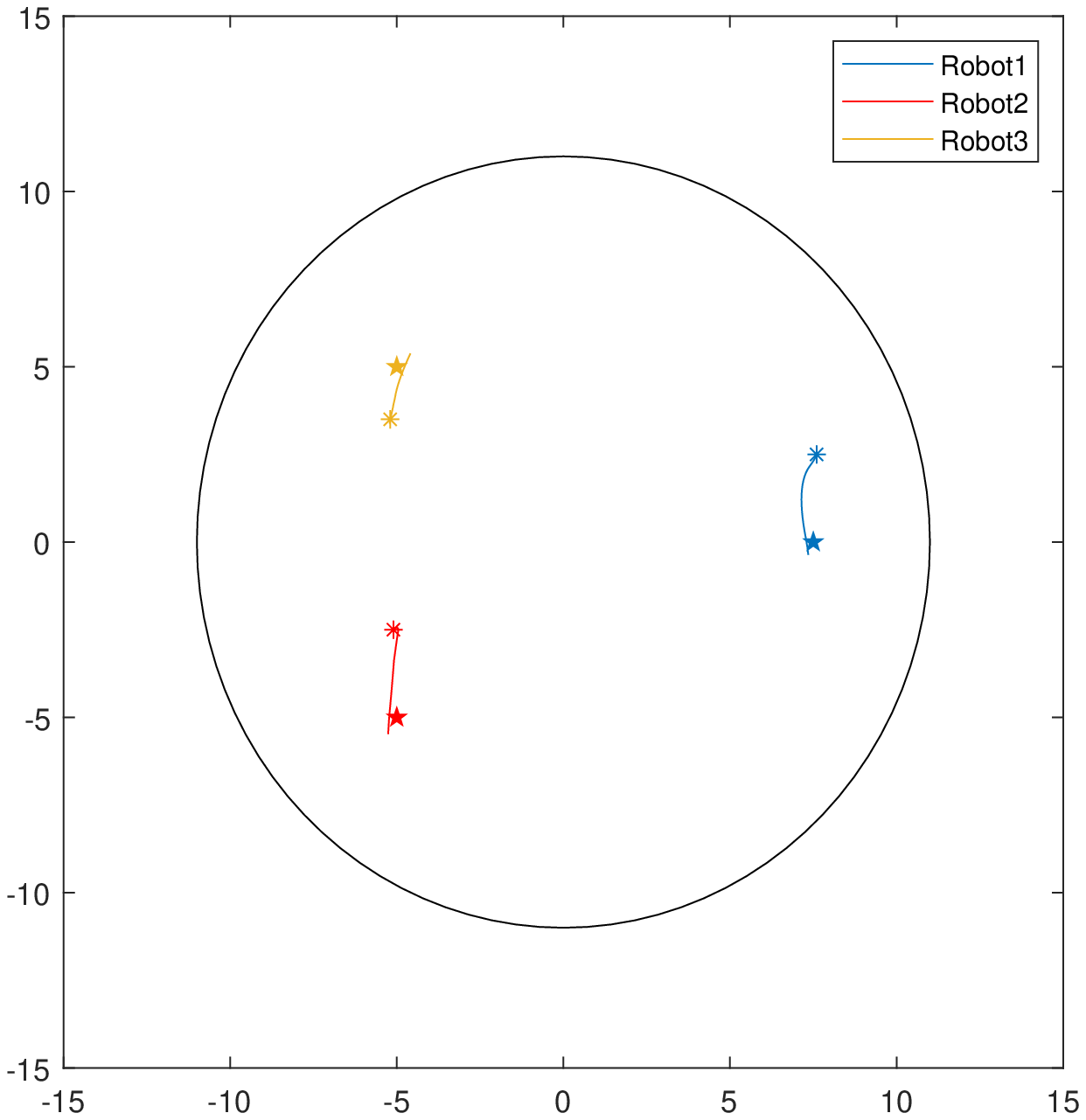}
         \caption{Case III (bilinear model based control design)}
         \label{Ctrl_eva_1_CaseIII}
     \end{subfigure}
     \hfill
     \begin{subfigure}[b]{0.48\textwidth}
         \centering
         \includegraphics[width=0.75\textwidth]{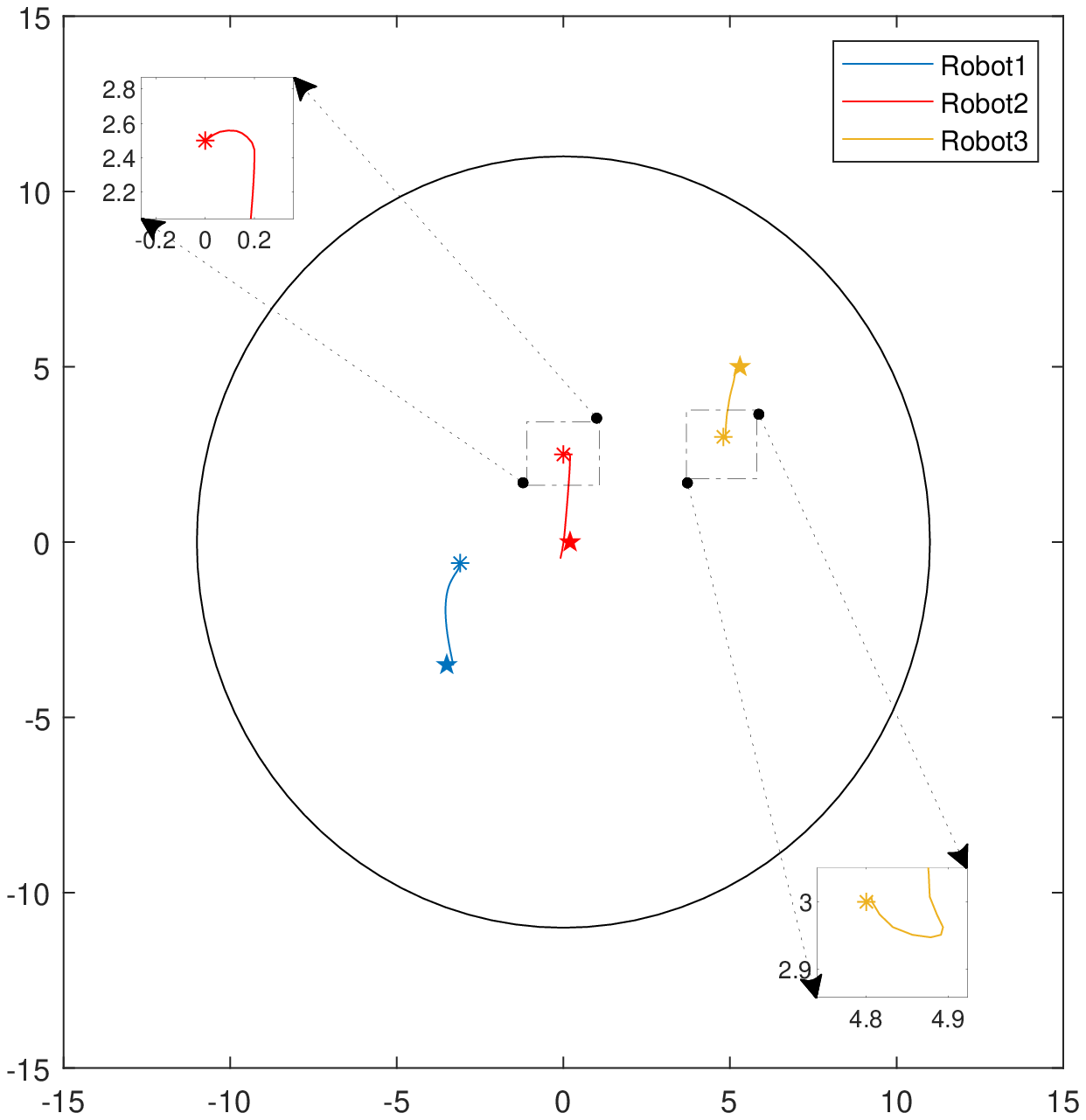}
         \caption{Case IV (bilinear model based control design)}
         \label{Ctrl_eva_1_CaseIV}
     \end{subfigure}
     \begin{subfigure}[b]{0.48\textwidth}
         \centering
         \includegraphics[width=0.75\textwidth]{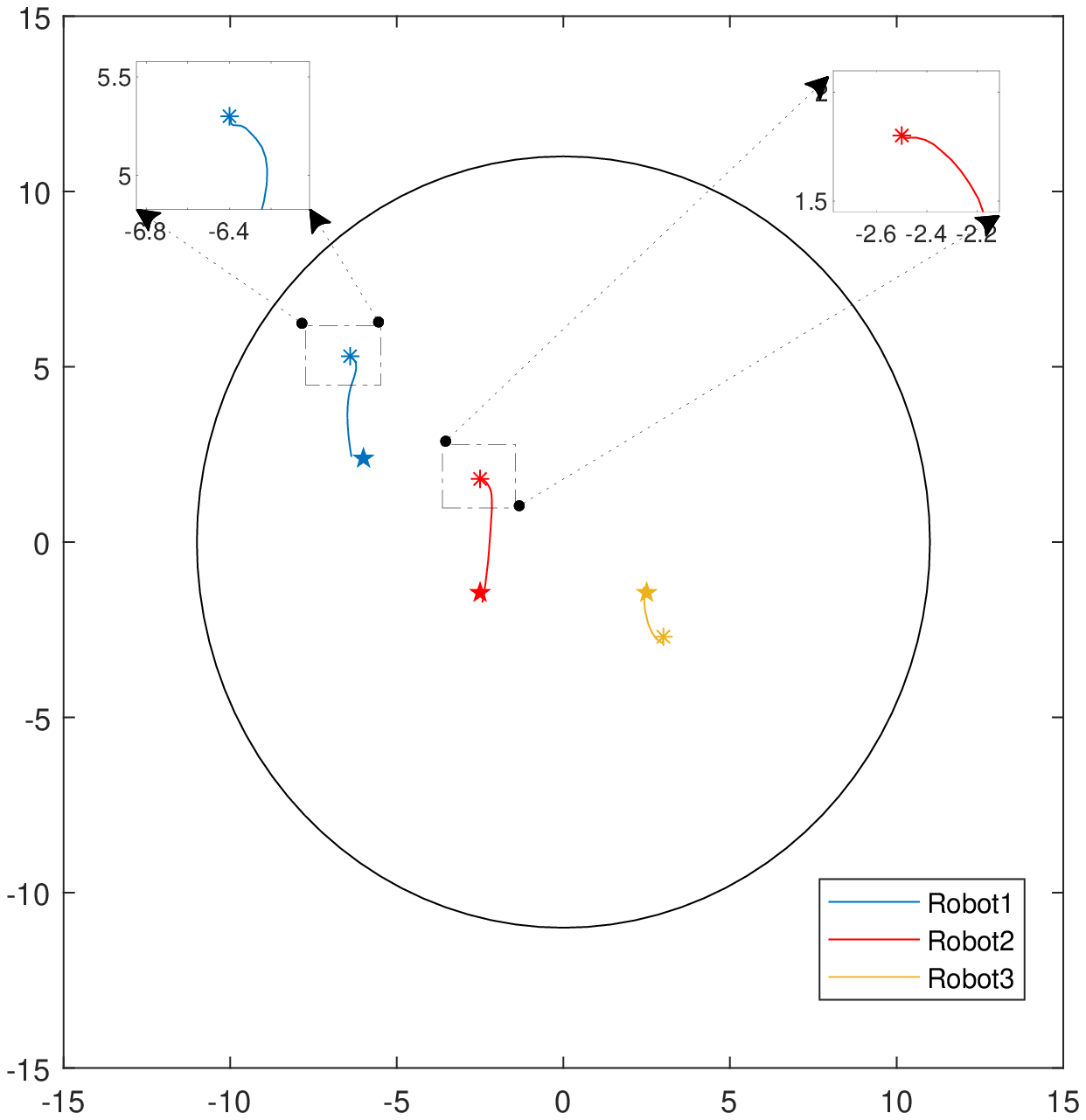}
         \caption{Case V (bilinear model based control design)}
         \label{Ctrl_eva_1_CaseV}
     \end{subfigure}

        
        \small
The asterisk markers $*$ in different colors indicate the initial positions of the robots, and the five-point-star markers $\star$ represent the target positions of the robots. The lines in different colors are the trajectories of the robots and the big black circle is the circular wall of the workspace for the robots. 
\caption{Robot trajectories in different cases.}
        \label{Ctrl_eva_1}
\end{figure}

The linear model based control design and the bilinear based control design simulation results are presented in Figure \ref{Ctrl_eva_1}. The plot of the robot trajectories controlled by the linear model based control design during the corresponding control evaluation phase is shown in Figure \ref{Ctrl_eva_1_Linear}. Then, Figures \ref{Ctrl_eva_1_CaseI}-\ref{Ctrl_eva_1_CaseV} show the robot trajectories controlled by the bilinear model based control design during the control evaluation phase in all five cases.

\bigskip
\textbf{Remark}. According to the analysis of the distances between the robots and the circular wall in all cases, the control evaluation is collision-free since all the distances are larger than zero indicating that the robots do not collide with others or with the wall in all five cases. 

\bigskip
{\bf Discussion}. From Figure \ref{Ctrl_eva_1}, only the bilinear model based control design can achieve the control objective to make all the robots arrive at their targets without collisions. According to Figure \ref{Ctrl_eva_1_Linear}, only Robot3 passes by the vicinal positions of its corresponding target. The other two robots cannot move toward the desired positions. Thus, although the linear model based control can achieve part of the control objective, a more elaborate approximation model should be considered to achieve the control objective. 

Figures \ref{Ctrl_eva_1_CaseI}-\ref{Ctrl_eva_1_CaseV} demonstrate the robots can reach the corresponding targets (or vicinal positions) simultaneously controlled by the baseline control input based on the estimated bilinear model for the $3-$robot system with the nonlinear utility functions in all five cases. This proves that the proposed optimal design based on the bilinear approximation model for the multi-robot system with the nonlinear utility functions can achieve the control objective which is reaching the target positions and collision-free.  

Additionally, in Figure \ref{Ctrl_eva_1_CaseIV}, the blue robot (Robot$1$) moves down to its left initially. The red robot (Robot$2$) moves toward our right-hand side direction at first. Then, it moves downward. The yellow robot (Robot$3$) also moves towards our right-hand side at first. This happens because the three robots interact with each other to avoid possible collisions since their initial distances are too short. Similarly, in Figure \ref{Ctrl_eva_1_CaseV}, we can see that both the blue robot (Robot$1$) and the red robot (Robot$2$) do not move directly toward their targets. Instead, the blue robot (Robot$1$) moves away from the wall (the lower right direction of their initial direction) at first to avoid the possible collision between itself and the wall. At the same time, the red robot (Robot$2$) moves toward our right-hand side at first to keep a safe distance from the approaching Robot$1$. The reason for this phenomenon is that the initial distance between the blue robot and the red robot and the initial distance between the blue robot and the wall are too close. These two cases show that the algorithm can make robots interact with each other to avoid possible collisions.


\par For the bilinear model based optimal control simulations of the selected five cases, the control evaluation phase lasts 30 iterations, which indicates 1.5 seconds in the actual situation based on the sampling time $\Delta t = 0.05$ seconds. The simulation program for the control evaluation phase costs 1.30 seconds for Case 1, 1.29 seconds for Case 2, 1.31 seconds for Case 3, 1.35 seconds for Case 4, and 1.33 seconds for Case 5 based on i7-8700k CPU. Thus, the ability of the control algorithm to solve the real problem is further proved. However, this control algorithm may not be global. In some cases, the robots' initial positions are far away from their targets, the robots cannot reach the target positions (or vicinal target positions).

\subsection{Simulations Results of Decentralized Control }
 \label{Sec_simuCEDec}
In this subsection, the simulation system and the simulation results for evaluating the bilinear model based decentralized control design of the 3-robot system are demonstrated sequentially. The robot trajectories are presented to check whether the desired tracking objectives are achieved. During the feedback control phase, the distances between robots with others and robots with the wall are analyzed to check whether collisions happen. 

\subsubsection{Simulation Conditions}
Similar to previous simulations, the simulation of the decentralized cases includes two phases too: \begin{itemize}
    \item \textbf{Phase I (System Identification Phase)}: Initially, we set the sampling time $\Delta t = 0.05 $s, $\rho=0.000 1$ as the iterative algorithm design parameter, $P_i(-1) = P_i^T(-1) = 100I_{109} >0, i = 1,2,3$ and $\Theta_i(0)=0_{109\times 6}$ as the initial values of the iterative algorithm for every single robot. The initial value of the robot state vectors are defined as $z_{(1),1}(0) = [x_1(0),y_1(0),v^x_1(0),v^y_1(0)]^T = [-8.5 ,-1, 0 ,0]^T $, $z_{(1),2}(0) = [x_2(0),y_2(0),v^x_2(0),v^y_2(0)]^T = [0, 8.5,\\ 0 ,0]^T $, $z_{(1),3}(0) = [x_3(0),y_3(0),v^x_3(0),v^y_3(0)]^T = [8.5, 0, 0,0]^T $ which means Robot 1, Robot 2, and Robot 3 are located at $(-8.5,-1)$, $(0,8.5)$, and $(8.5,0)$ with zero initial speeds in X-direction and Y-direction at $k=0$. Correspondingly, the target positions of the three robots are $(-6.5,0)$, $(0, 6.5)$, and $(6.5,0)$. In the first $k_s=300$ iterations, apply a selected proper input vector to do the linear or bilinear model approximation according to \cite{t21},\cite{zt22}. The detailed steps for each system identification iteration are listed in the following:
    \begin{enumerate}
        \item [\textbf{Step 1}: ] Compute the input $U_i(k)$, whose components are listed in Table \ref{S1_MI_in}. 
        \item [\textbf{Step 2}: ] Update all the vectors for every robot $z_{(1),i}(k), i = 1,2,3$, according to (\ref{z1_def_d}). 
\item [\textbf{Step 3}: ] Calculate all the vectors for every robot $z_{(2),i}(k), i = 1,2,3$, based on equation \eqref{z2_def_d}.

\item [\textbf{Step 4}: ] Update the approximations for each robot $\hat{z}_{(2),i}(k+1) = \hat{\Theta}_i^T(k)\zeta_i(k), i = 1,2,3$, with $\hat{\Theta}^T_i(k) $, $\zeta_i(k) $ for the nonlinear utility functions based on equations (\ref{epsilon(k)})-(\ref{Pk1}). The signal vector $\zeta_i(k) $ and the parameter matrix $\hat{\Theta}_i(k) $ are defined in equations (\ref{zeta_bilinear_d}) and (\ref{theta_hat_bilinear_d}). 
    \end{enumerate}

    \item \textbf{Phase II (Feedback Control Phase)}: From the $(k_s+1) = 301 $th iteration, reset the initial and target positions of the robots and replace the input vector with the results of the linear programming problem to test the baseline control input performance and keep the updating of the model approximation results. Table \ref{robot_setting_bi_d} demonstrates the detailed robot initial and target position settings of the selected two cases. The details are listed in the following:

 \begin{enumerate}
        \item [\textbf{Step 1}: ]Generate the input $U_i(k)$, which is the solution for the linear programming problem shown in equation (\ref{sol_inputfrombilinear_d}). 
        
       \item [\textbf{Step 2}: ] Update all the vectors for every robot $z_{(1),i}(k), i = 1,2,3$, according to (\ref{z1_def_d}). 
\item [\textbf{Step 3}: ] Calculate all the vectors for every robot $z_{(2),i}(k), i = 1,2,3$, based on equation \eqref{z2_def_d}.

\item [\textbf{Step 4}: ] Update the approximations for each robot $\hat{z}_{(2),i}(k+1) = \hat{\Theta}_i^T(k)\zeta_i(k), i = 1,2,3$, with $\hat{\Theta}^T_i(k) $, $\zeta_i(k) $ for the nonlinear utility functions based on equations (\ref{epsilon(k)})-(\ref{Pk1}). The signal vector $\zeta_i(k) $ and the parameter matrix $\hat{\Theta}_i(k) $ are defined in equations (\ref{zeta_bilinear_d}) and (\ref{theta_hat_bilinear_d}).

    \end{enumerate}
\begin{table}[H]
    \renewcommand\arraystretch{1.5}

\begin{center}
\begin{tabular}{cccc} 
\hline
\hline
\multicolumn{4}{c}{Case Settings}\\
\hline
Case number             &  Robot number             & Initial position & Target position\\
\hline
\multirow{3}{*}{Case I}              
                        &\multirow{1}{*}{Robot 1}   &      $(-2.5,7.1)$  & $(-3.5, 5)$     \\ 
                        &\multirow{1}{*}{Robot 2}   &     $(2.7,3.3)$ & $(3.5 ,5)$     \\ 
                        &\multirow{1}{*}{Robot 3}   &     $(-0.8,-3.5)$  & $(0 ,-5.5)$   \\

\hline

\multirow{3}{*}{Case II}              
                        &\multirow{1}{*}{Robot 1}   &      $(-6.5,-2.9)$&$(-4.5,0)$    \\ 
                        &\multirow{1}{*}{Robot 2}   &     $(1.7,2.3)$&$(3,4)$     \\ 
                        &\multirow{1}{*}{Robot 3}   &     $(2.4,-5.2)$&$(3,-4)$   \\ 
                        \hline\hline
\end{tabular}
\end{center}
\caption{Settings of the feedback control phase.}

\label{robot_setting_bi_d}
\end{table}    
    
\end{itemize}

\subsubsection{Simulation Results}
\begin{figure}[H]
     \centering
     \begin{subfigure}[b]{0.4\textwidth}
         \centering
         \includegraphics[width=\textwidth]{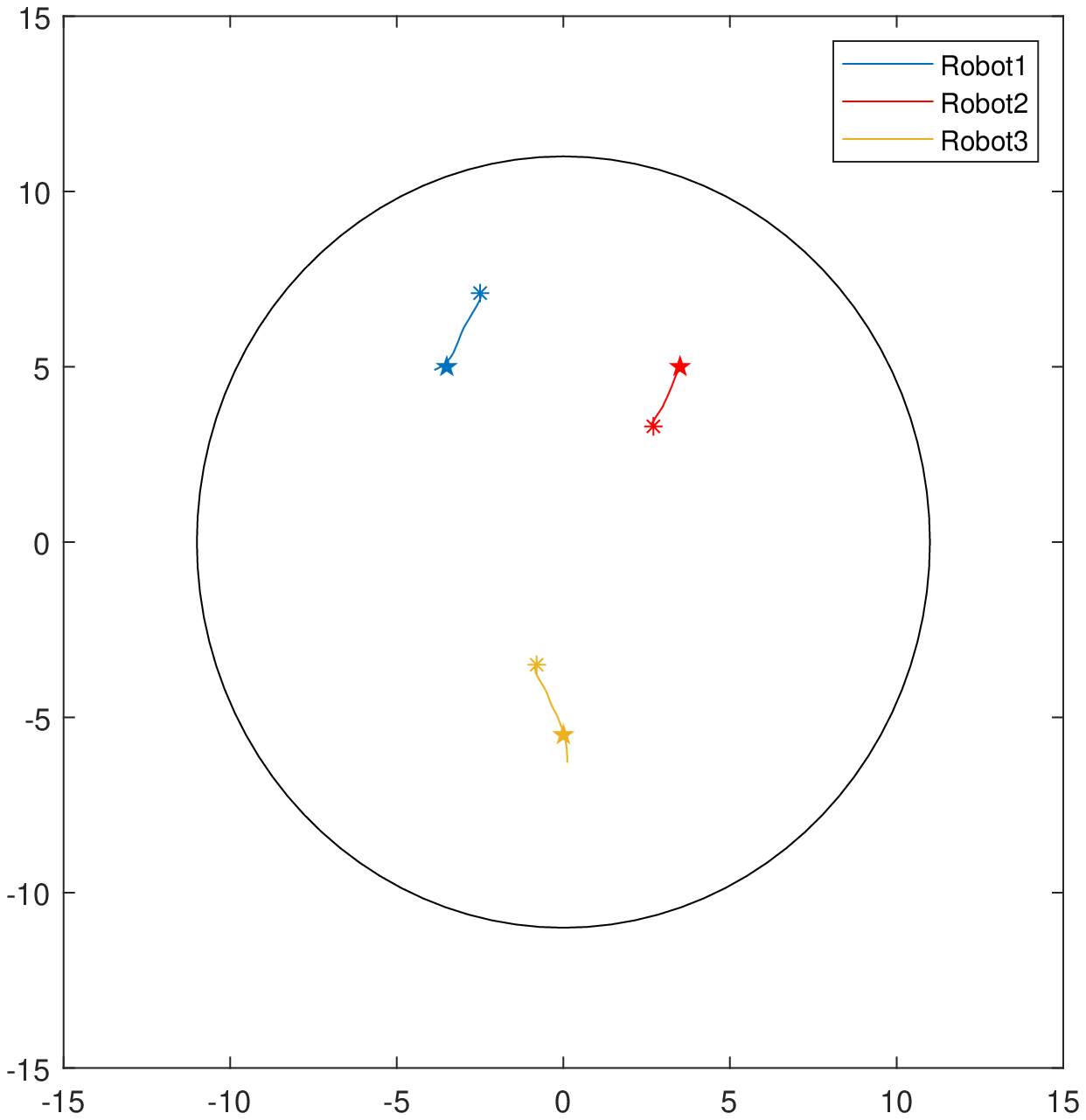}
         \caption{Case I }
         \label{Ctrl_eva_d1}
     \end{subfigure}
     \hfill
     \begin{subfigure}[b]{0.4\textwidth}
         \centering
         \includegraphics[width=\textwidth]{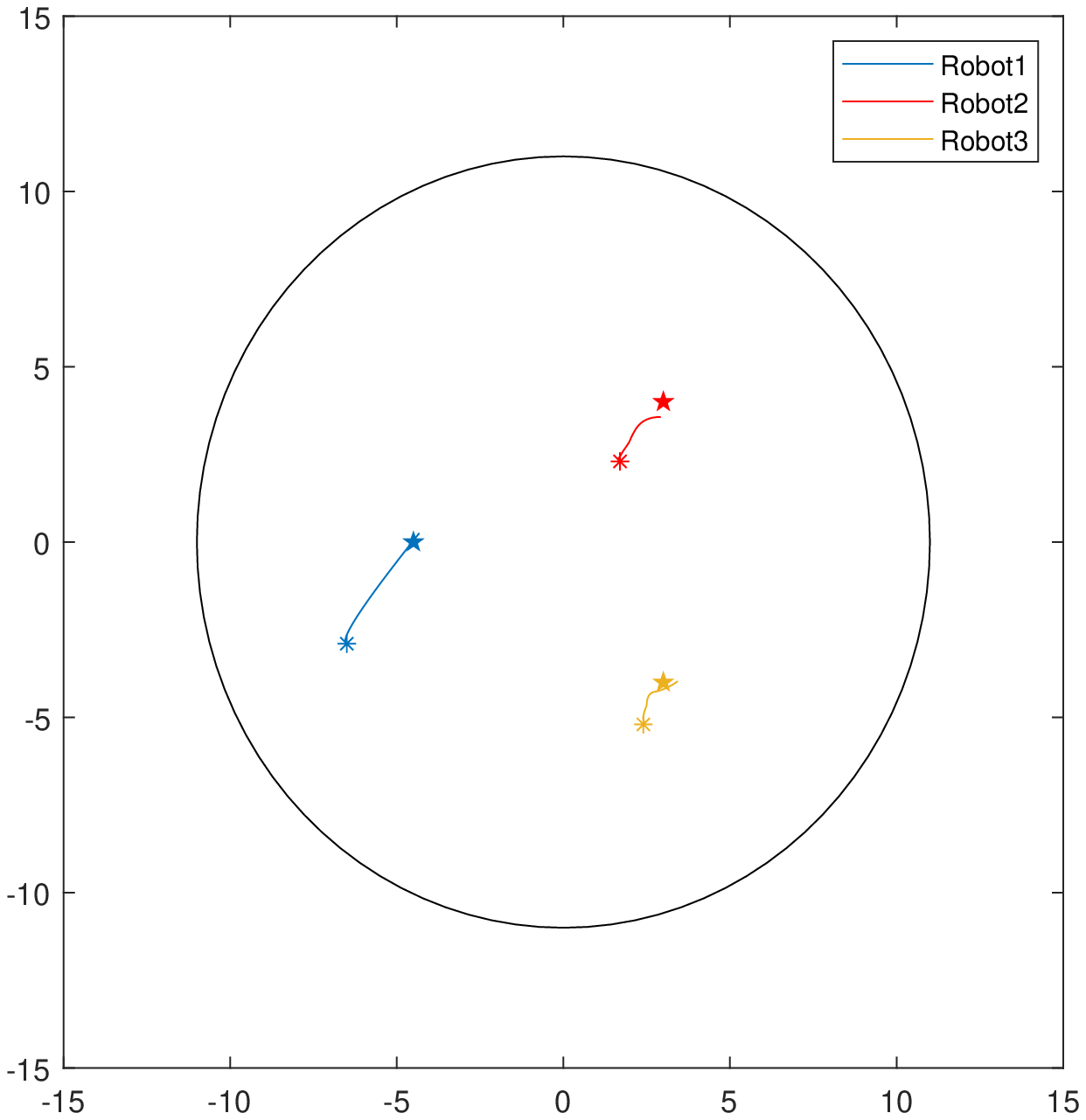}
         \caption{Case II}
         \label{Ctrl_eva_d2}
     \end{subfigure}
     \hfill
     \medskip
        
        \small
The asterisk markers $*$ in different colors indicate the initial positions of the robots, and the five-point-star markers $\star$ represent the target positions of the robots. The lines in different colors are the trajectories of the robots and the big black circle is the circular wall of the workspace for the robots. 
\caption{Robot trajectories in different cases (Bilinear model based decentralized control).}
        \label{Ctrl_eva_d}
\end{figure}

The bilinear based decentralized control design simulation results are presented in Figure \ref{Ctrl_eva_d}.

\bigskip
{\bf Model identification results}. The results of the model identification phases are almost the same as the results in Section 5.2. The \textit{posteriori} estimation error in the case applying the decentralized bilinear model identification algorithm is always small without any periodical changes.

\bigskip
\textbf{Optimal control results}.
In Figure \ref{Ctrl_eva_d}, (a)-(b) demonstrate the robots can reach the corresponding targets (or vicinal positions) simultaneously controlled by the decentralized generated control input based on the estimated bilinear model for the 3 robot system with the nonlinear utility functions in both cases. This proves that the decentralized version optimal design based on the bilinear approximation model for the multi-robot system with the nonlinear utility functions can achieve the control objective which is reaching the target positions and collision-free.  

\par For the simulations of the selected two cases, the control evaluation phase lasts 30 iterations, which indicates 1.5 seconds in the actual situation based on the sampling time $\Delta t = 0.05$ seconds. The simulation program of the control evaluation phase for the three robots costs 1.73 seconds for Case I, and 1.71 seconds for Case II based on i7-8700k CPU. Since the control inputs for each robot are generated decentralized, we can use three computers to find the optimal control input simultaneously. Thus, the running time of the control evaluation phase for every single robot can be decreased to one-third of the listed running time for all three robots. This proves that not only the decentralized bilinear model based control design is fast enough to be implemented for real-time control, but also that the algorithm can be accelerated by applying its decentralized version with more computation resources. 

\subsection{Additional Parameter Estimation Algorithm: Normalized Adaptive Gradient Algorithm}

The least-squares algorithm \eqref{Thetak+1}-\eqref{Pk1} has a big matrix $P(k)$ to calculate. We also simulated the cases that applied another parameter estimation algorithm called the normalized adaptive gradient algorithm to see whether it can make the results better.

With the same definition of the estimation error $\epsilon(k) = \hat{\Theta}^T(k)\zeta(k)-y(k)$ as the least squares algorithm and the measured vector signals $y(k)\in R^{18}$ and $\zeta(k) \in R^{q}$ \footnote{For the linear model, the value of $q$ is 36. For the bilinear model, the value of $q$ is 901.}, the adaptive law of the gradient algorithm is defined as 
\begin{equation}
    \hat{\Theta}(k+1) = \hat{\Theta}(k) - \frac{\Gamma\zeta(k)\epsilon^T(k)}{m^2(k)}, k = k_0,\,k_0+1,\,k_0+2,\cdots, 
\end{equation}
where $0<\Gamma=\Gamma^T<2I_q$ \footnote{The matrix $I_q$ denotes the identity matrix of size $q\times q$. }, and 
\begin{equation}
    m(k) = \sqrt{\rho+\zeta^T(k) \zeta(k)},\,\rho >0.  
\end{equation}

\subsubsection{Simulation Results}
In the simulation, We replace the least squares parameter estimation algorithm of Step 4 in both of the system identification phase and the feedback control phase of the bilinear model based control evaluation simulation procedures in Section \ref{Sec_simuCE} with the gradient algorithm. 

With the application of the gradient algorithm as  parameter estimation method, the robot trajectories in the simulation results for all the cases in Section \ref{Sec_simuCE} show the robots cannot reach their targets. Although different values of $\Gamma$ lead to different trajectories, most trajectories are even almost straight lines without any collision avoidance behaviors or moving direction adjustment toward the corresponding target behaviors.

\section{Concluding Remarks}
In this report, we have finished the simulation studies of the Koopman system approximation based optimal control of multiple robots with nonlinear utility functions, which is detailed in \cite{tzh22a}. The desired robot tracking performance is ensured by maximizing the nonlinear utility function. In \cite{tzh22a}, an algorithm to find the linear or bilinear approximations of the nonlinear utility function components and the optimal control design based on the estimated linear or bilinear approximations. In this report, the simulation results of the linear or bilinear approximation algorithm and the control designs based on the approximations are reported and analyzed. 
 
 Specifically, we formulate a 3-robot system to examine the performance of the control designs based on the linear approximation model and bilinear approximation model. According to the linear and bilinear model identification simulation results in Section \ref{Sec_simuLinID}-\ref{Sec_simuBilinID}, only the centralized bilinear model can approximate the nonlinear utility functions well since only the bilinear model identification can ensure the \textit{posteriori} estimation error within a smaller range, i.e. $|\epsilon_a| = (0,3\times10^{-4})$. 
 Simulation results in Section \ref{Sec_simuCE} show that the centralized linear model based control design can only achieve parts of the tracking objectives, which means the linear model does not contain rich enough information to construct the optimal control design. After that, other results in Section \ref{Sec_simuCE} prove that the centralized bilinear model based optimal control design can achieve the desired tracking objectives in all three listed cases under the computation time restriction for the real-time control. Lastly, Section \ref{Sec_simuCEDec} shows the decentralized bilinear model based control signals can also achieve the desired tracking objectives faster. The success of the simulation for the decentralized bilinear model based control inputs infers the proposed bilinear model based optimal control algorithm owns the potential to be applied to more complex applications, like autonomous driving. 

 In the future, the performance of the bilinear model based optimal control design can be improved to ensure the desired global tracking performance within a shorter computation time. 
\section*{Acknowledgements}
The authors would like to thank the financial support from a Ford University Research Program
grant and the collaboration and help from Dr. Suzhou Huang and Dr. Qi Dai of Ford Motor
Company for this research.

\end{document}